\documentclass[preprint]{aastex}

\usepackage{pdflscape}
\usepackage{paralist}
\usepackage{amsmath}
\usepackage{epsf}
\usepackage{epstopdf}
\usepackage{graphicx}

\shorttitle{Surface gravity limits \\
from non-detection of oscillations}
\shortauthors{Campante et al.}

\begin{document}

\title{Limits on surface gravities of {\it Kepler} planet-candidate host stars \\
from non-detection of solar-like oscillations}

\author{T.~L.~Campante\altaffilmark{1}}
\email{campante@bison.ph.bham.ac.uk}
\and
\author{
    W.~J.~Chaplin\altaffilmark{1},
    M.~N.~Lund\altaffilmark{2},
    D.~Huber\altaffilmark{3,4},
    S.~Hekker\altaffilmark{5,6},
    R.~A.~Garc\'ia\altaffilmark{7},
    E.~Corsaro\altaffilmark{8,9},
    R.~Handberg\altaffilmark{1},
    A.~Miglio\altaffilmark{1},
    T.~Arentoft\altaffilmark{2},      
    S.~Basu\altaffilmark{10},
    T.~R.~Bedding\altaffilmark{11},
    J.~Christensen-Dalsgaard\altaffilmark{2},
    G.~R.~Davies\altaffilmark{1},
    Y.~P.~Elsworth\altaffilmark{1},
    R.~L.~Gilliland\altaffilmark{12},    
    C.~Karoff\altaffilmark{2},
    S.~D.~Kawaler\altaffilmark{13},
    H.~Kjeldsen\altaffilmark{2},
    M.~Lundkvist\altaffilmark{2},
    T.~S.~Metcalfe\altaffilmark{14,2},
    V.~Silva Aguirre\altaffilmark{2},
    D.~Stello\altaffilmark{11}
}

\altaffiltext{1}{School of Physics and Astronomy, University of
   Birmingham, Edgbaston, Birmingham, B15 2TT, UK}

\altaffiltext{2}{Stellar Astrophysics Centre (SAC), Department of
   Physics and Astronomy, Aarhus University, Ny Munkegade 120, DK-8000
   Aarhus C, Denmark}
   
\altaffiltext{3}{NASA Ames Research Center, MS 244-30, Moffett Field,
   CA, 94035, USA}

\altaffiltext{4}{NASA Postdoctoral Program Fellow}
   
\altaffiltext{5}{Astronomical Institute, ``Anton Pannekoek'',
   University of Amsterdam, The Netherlands}
   
\altaffiltext{6}{Max Planck Institute for Solar System Research, 
   Katlenburg-Lindau, Germany}      

\altaffiltext{7}{Laboratoire AIM, CEA/DSM-CNRS-Universit\'e Paris Diderot; IRFU/SAp, Centre de Saclay, 
   91191 Gif-sur-Yvette Cedex, France}

\altaffiltext{8}{Instituut voor Sterrenkunde, KU Leuven, Celestijnenlaan 200D, B-3001 Leuven, Belgium}

\altaffiltext{9}{INAF --- Astrophysical Observatory of Catania, Via S. Sofia 78, I-95123 Catania, Italy}

\altaffiltext{10}{Department and Astronomy, Yale University, New Haven,
   CT, 06520, USA}

\altaffiltext{11}{Sydney Institute for Astronomy, School of Physics,
   University of Sydney, Sydney, Australia}

\altaffiltext{12}{Center for Exoplanets and Habitable Worlds, The
   Pennsylvania State University, University Park, PA, 16802, USA}

\altaffiltext{13}{Department of Physics and Astronomy, Iowa State
   University, Ames, IA, 50011, USA}

\altaffiltext{14}{Space Science Institute, Boulder, 
   CO, 80301, USA}

\begin{abstract}
We present a novel method for estimating lower-limit surface gravities ($\log g$) of {\it Kepler} targets whose data do not allow the detection of solar-like oscillations. The method is tested using an ensemble of solar-type stars observed in the context of the {\it Kepler} Asteroseismic Science Consortium. We then proceed to estimate lower-limit $\log g$ for a cohort of {\it Kepler} solar-type planet-candidate host stars with no detected oscillations. Limits on fundamental stellar properties, as provided by this work, are likely to be useful in the characterization of the corresponding candidate planetary systems. Furthermore, an important byproduct of the current work is the confirmation that amplitudes of solar-like oscillations are suppressed in stars with increased levels of surface magnetic activity.
\end{abstract}


\section{Introduction}\label{sec:intro}
The NASA {\it Kepler} mission was designed to use the transit method to detect Earth-like planets in and near the habitable zones of late-type main-sequence stars \citep{KeplerMission,KeplerDesign}. {\it Kepler} has yielded several thousands of new exoplanet candidates \citep{BoruckiCandidates1,BoruckiCandidates2,BatalhaCandidates3}, bringing us closer to one of the mission's objectives, namely, the determination of planet occurrence rate as a function of planet radius and orbital period. However, indirect detection techniques, such as transit and radial velocity observations, are only capable of providing planetary properties relative to the properties of the host star. Therefore, accurate knowledge of the fundamental properties of host stars is needed to make robust inference on the properties of their planetary companions. 

Unfortunately, the vast majority of the planet-candidate host stars -- also designated as {\it Kepler} Objects of Interest or KOIs -- are too faint to have measured trigonometric parallaxes, so that most of the currently available stellar parameters rely on a combination of ground-based multicolor photometry, spectroscopy, stellar model atmospheres and evolutionary tracks. This is the case for the {\it Kepler} Input Catalog \citep[KIC;][]{KIC}. Based on an asteroseismic analysis, \citet{VernerKIC} have detected an average overestimation bias of $0.23\:{\rm dex}$ in the KIC determination of the surface gravity for stars with $\log g_{\rm KIC}\!>\!4.0\:{\rm dex}$, thus implying an underestimation bias of up to $50\,\%$ in the KIC radii for stars with $R_{\rm KIC}\!<\!2\,{\rm R_\sun}$ \citep[see also][]{BrunttSpec}. \citet{KIC} had flagged this behavior, warning that the KIC classifications tend to give $\log g$ too large for subgiants. This is a natural cause for concern if these values are to be used in the characterization of exoplanetary systems. This situation can be improved for stellar hosts for which high-resolution spectroscopy is available, an example being the metallicity study undertaken by \citet{Buchhave} on a sample of F, G and K dwarfs hosting small exoplanet candidates. However, spectroscopic methods are known to suffer from degeneracies between the effective temperature $T_{\rm eff}$, the iron abundance $[\rm{Fe}/\rm{H}]$ and $\log g$, yielding constraints on the stellar mass and radius that are model-dependent \citep[e.g.,][]{Torres}. The planet-candidate catalog provided by \citet{BatalhaCandidates3}, based on the analysis of the first 16 months of data (from quarter Q1 to quarter Q6), includes a valuable revision of stellar properties based on matching available constraints (from spectroscopic solutions, whenever available, otherwise from the KIC) to Yonsei-Yale evolutionary tracks \citep{Demarque}.

Asteroseismology can play an important role in the determination of accurate fundamental properties of host stars. Solar-like oscillations in a few tens of main-sequence stars and subgiants have been detected using ground-based high-precision spectroscopy \citep[e.g.,][]{BouchyCarrier,18Sco} and ultra-high-precision, wide-field photometry from the {\it CoRoT} space telescope \citep[e.g.,][]{Appour08,MichelSci}. {\it Kepler} photometry has ever since revolutionized the field of solar-like oscillations by leading to an increase of one order of magnitude in the number of such stars with confirmed oscillations \citep[][]{VernerComparison}. In particular, {\it Kepler} short-cadence data \citep[$\Delta t\!\sim\!1\:{\rm min}$;][]{GillilandNoise} make it possible to investigate solar-like oscillations in main-sequence stars and subgiants, whose dominant periods are of the order of several minutes. The information contained in solar-like oscillations allows fundamental stellar properties (i.e., density, surface gravity, mass and radius) to be determined \citep[e.g.,][and references therein]{ChaMigReview}. The very first seismic studies of exoplanet-host stars were conducted using ground-based \citep[][]{Bouchy05,Vauclair08} and {\it CoRoT} data \citep[][]{Gaulme10,Ballot11}. \citet{JCDExo} reported the first application of asteroseismology to known exoplanet-host stars in the {\it Kepler} field. Subsequently, asteroseismology has been used to constrain the properties of {\it Kepler} host stars in a series of planet discoveries \citep{Kepler10,TrES2,Kepler22,Kepler36,Kepler21,Kepler69,ChaplinObliquities,Kepler68,Misalignment,Kepler410A}. Recently, the first systematic study of {\it Kepler} planet-candidate host stars using asteroseismology was presented by \citet{HuberEnsembleKOI}.

In this work, we present a novel method for placing limits on the seismic and thus fundamental properties of {\it Kepler} targets whose data do not allow to detect solar-like oscillations (Sect.~\ref{sec:method}). Our method relies on being able to predict, for a data set of given stellar and instrumental noise, the threshold oscillation amplitude required to make a marginal detection of the oscillations. This threshold amplitude is frequency-dependent, as we shall explain in detail below. Moreover, on the basis of {\it Kepler} observations we determine the dependence of the maximum mode amplitude of solar-like oscillations on the frequency at which it is attained. As we shall see, this gives a well-defined amplitude trend. By comparing this trend to the frequency-dependent amplitude for marginal detection of the oscillations, we may set limits on the seismic parameters and hence stellar properties that are required for marginal detection. In Sect.~\ref{sec:resosc}, the method is tested using an ensemble of solar-type stars observed in the context of the {\it Kepler} Asteroseismic Science Consortium \citep[KASC;][]{KAI,KAI2}. Finally, lower-limit $\log g$ estimates are provided for a cohort of {\it Kepler} solar-type planet-candidate host stars with no detected oscillations (Sect.~\ref{sec:resnonosc}). We discuss the potential use and the limitations of our work in Sect.~\ref{sec:discussion}.

\section{Method description}\label{sec:method}
\subsection{Overview}\label{sec:overview}
Solar-like oscillations are predominantly global standing acoustic waves. These are p modes (the pressure variation playing the role of the restoring force) and are characterized by being intrinsically damped while simultaneously stochastically excited by near-surface convection \citep[e.g.,][]{JCDReview}. Therefore, all stars cool enough to harbor an outer convective envelope may be expected to exhibit solar-like oscillations. In the remainder of this work, the term {\it solar-type star} will be used to designate a wide range of F, G and K dwarfs and subgiants.

The frequency-power spectrum of the oscillations in solar-type stars and red giants presents a pattern of peaks with near-regular frequency separations \citep{Vandakurov,Tassoul}. The most prominent separation is the so-called large frequency separation, $\Delta\nu$, between neighboring overtones having the same spherical angular degree, $l$. Oscillation mode power is modulated by an envelope that is generally Gaussian-like in shape \citep[e.g.,][]{KallingerGaussEnvelope}. The frequency of the peak of the power envelope of the oscillations, where the observed modes present their strongest amplitudes, is commonly referred to as the frequency of maximum amplitude, $\nu_{\rm max}$. The maximum height (power spectral density), $H_{\rm max}$, of the envelope, and thus the maximum mode amplitude, $A_{\rm max}$, are strong functions of $\nu_{\rm max}$ \citep[e.g.,][]{Mosser12}.

\begin{figure}
\epsscale{1.2}
\plottwo{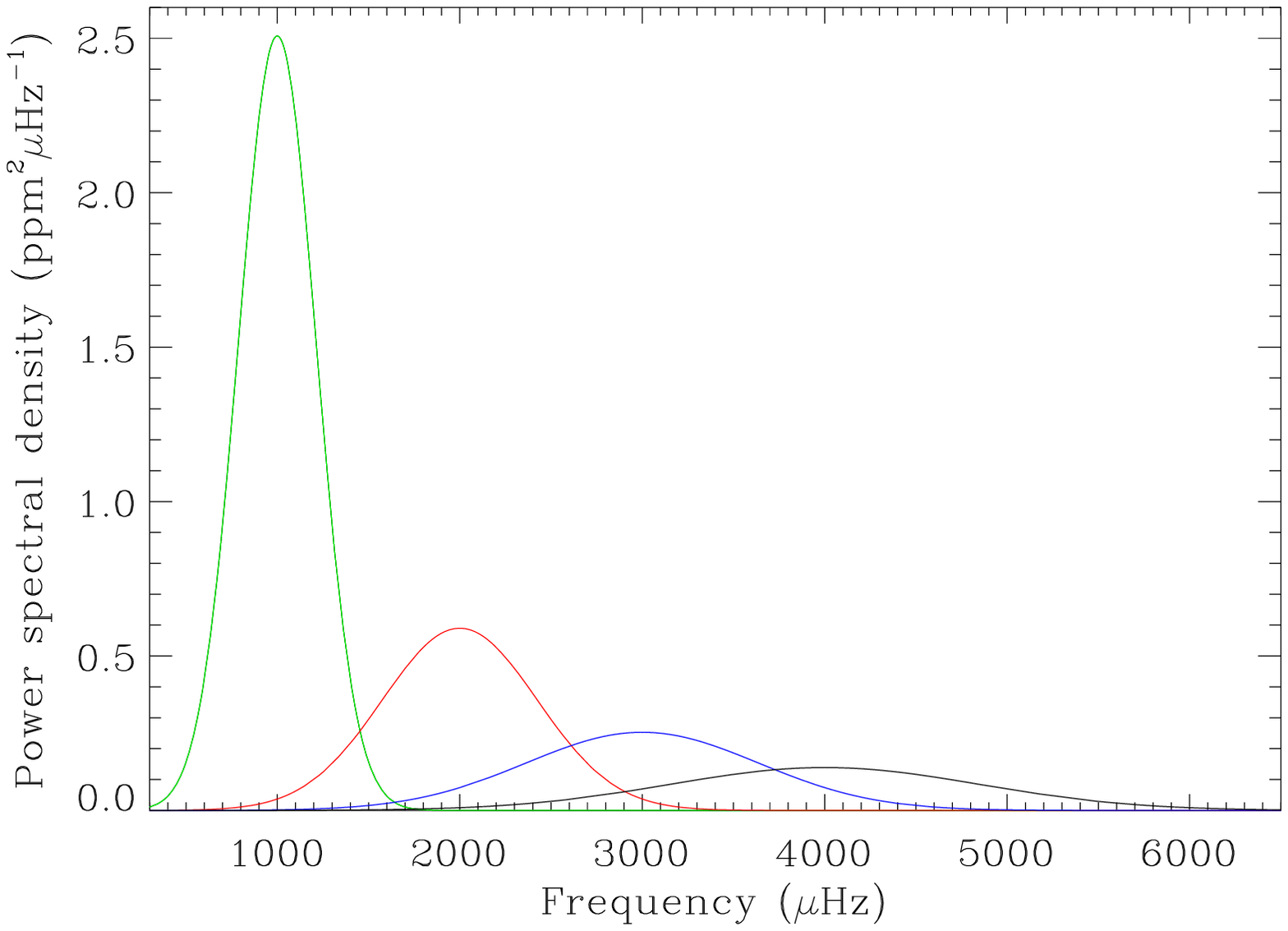}{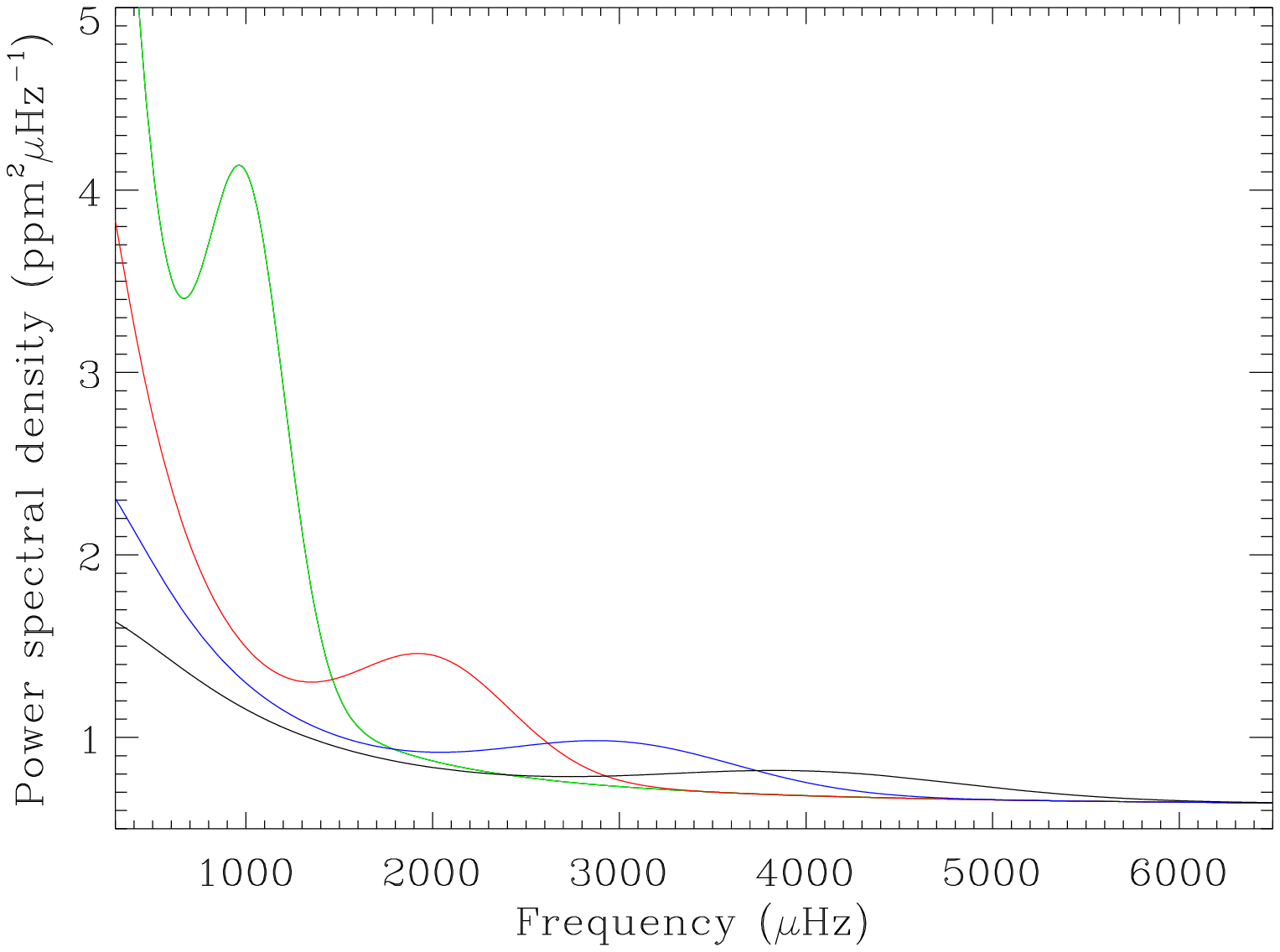}
\caption{\small Left-hand panel: Idealized oscillation power envelopes for stars showing solar-like oscillations with (from left to right) $\nu_{\rm max}\!=\!1000$, 2000, 3000 and $4000\:\rm \mu Hz$. Right-hand panel: Resulting idealized limit frequency-power spectra, from combination of oscillation power envelopes and expected background power-spectral densities from granulation and instrumental/shot noise.\label{fig:expow}}
\end{figure}

Figure \ref{fig:expow} shows some examples of idealized limit frequency-power spectra for stars displaying solar-like oscillations. The left-hand panel shows idealized oscillation power envelopes with $\nu_{\rm max}$ ranging from 1000 to $4000\:\rm \mu Hz$. As $\nu_{\rm max}$ decreases, the heights, $H_{\rm max}$, at the center of the power envelopes increase, and the envelopes also get narrower in frequency, with the FWHM being approximately given by $\nu_{\rm max}/2$ \citep[e.g.,][]{StelloM67,MosserRGCoRoT}, implying that most of the mode power is constrained to a range $\pm\nu_{\rm max}/2$ around $\nu_{\rm max}$. The right-hand panel shows the result of adding the expected limit background power-spectral density from granulation and instrumental/shot noise. The latter contribution, seen as a constant offset at high frequencies, depends on the stellar magnitude and has been computed following the empirical minimal term model for the noise given in \citet{GillilandNoise}, where one has assumed observations at {\it Kepler}-band magnitude $m_{\rm Kep}\!=\!9$. As $\nu_{\rm max}$ decreases, the power from granulation, modeled as a Lorentzian function centered on zero frequency \citep[e.g.,][]{Harvey}, is seen to increase while becoming more concentrated at lower frequencies \citep[][]{RGgran,Samadi13}.


Given the noise background and the length of the observations, one may estimate the oscillation amplitudes that are required to make a marginal detection against that background. To that end, we use the method described in \citet{ChaplinDetect}. Since the backgrounds presented by stars vary with frequency, the detection test must be applied at different frequencies within a comprehensive frequency range. At each frequency we estimate the signal-to-noise ratio, and hence the mode envelope height and maximum amplitude, needed to detect a spectrum of solar-like oscillations centered on that frequency (see Appendix \ref{sec:threshold} for details).


\begin{figure}
\epsscale{1.2}
\plottwo{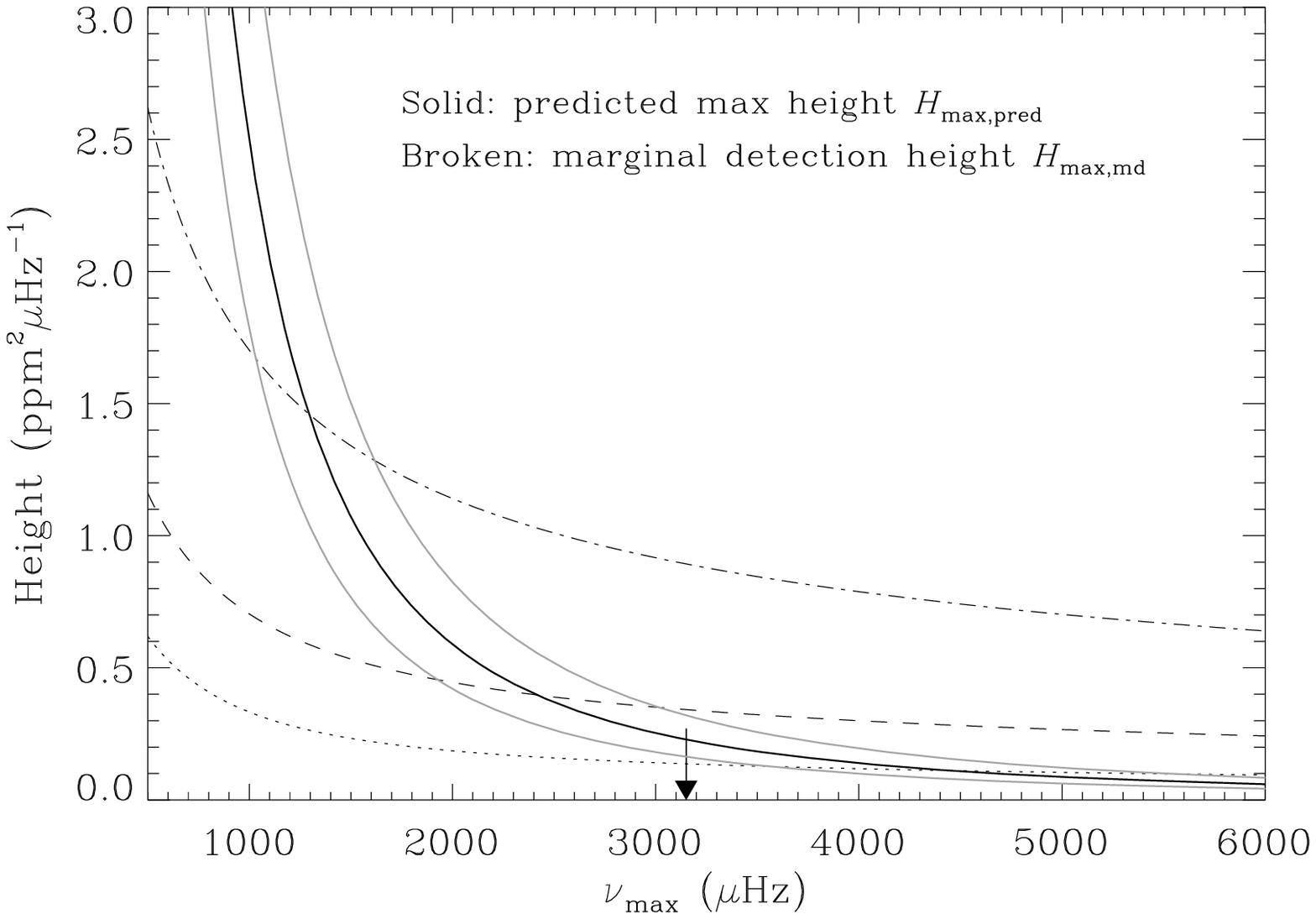}{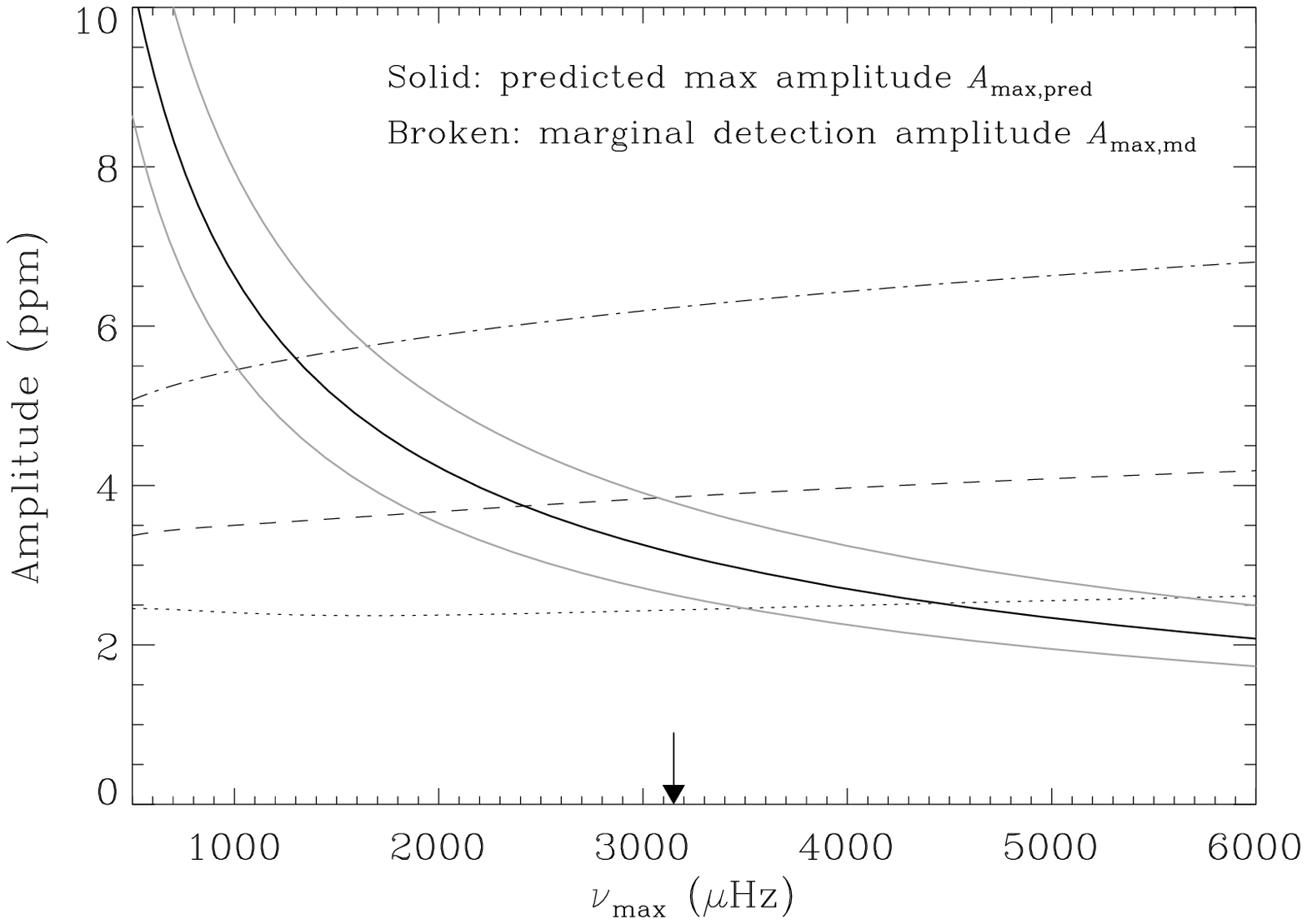}
\caption{\small Left-hand panel: Predicted trend in the mode power envelope height, $H_{\rm max,pred}$, as a function of $\nu_{\rm max}$ (solid black line). Solid gray lines represent the $\pm 1\sigma$ confidence interval on $H_{\rm max,pred}$. The frequency-dependent marginal height for detection, $H_{\rm max,md}$, is also shown assuming 1-month-long observations of a hypothetical solar twin made at {\it Kepler}-band magnitudes $m_{\rm Kep}\!=\!10$ (dotted), $m_{\rm Kep}\!=\!11$ (dashed) and $m_{\rm Kep}\!=\!12$ (dot-dashed). The arrow indicates $\nu_{\rm max,\sun}$. Right-hand panel: Same as left-hand panel, but for the amplitudes.\label{fig:exdetect}}
\end{figure}


The left-hand panel of Fig.~\ref{fig:exdetect} shows the threshold oscillation power envelope heights, $H_{\rm max,md}$, required for marginal detection of oscillations in a hypothetical solar twin. The calculations assumed 1-month-long observations at {\it Kepler}-band magnitudes $m_{\rm Kep}\!=\!10$ (dotted), $m_{\rm Kep}\!=\!11$ (dashed) and $m_{\rm Kep}\!=\!12$ (dot-dashed). As the noise level rises -- here, with increasing value of $m_{\rm Kep}$ -- so too do the threshold heights needed for detection. These thresholds also increase with decreasing frequency because the rising backgrounds make it potentially harder to detect the oscillations.

The solid black line shows the expected trend in height, $H_{\rm max,pred}$, as a function of $\nu_{\rm max}$ (as per the left-hand panel of Fig.~\ref{fig:expow}), the surrounding lines following the $\pm 1\sigma$ envelope given by analysis of {\it Kepler} data. This will be discussed in more detail in Sect.~\ref{sec:calibration}. We note that while this trend is a strong function of $\nu_{\rm max}$, allowance must also be made for some dependence on the effective temperature $T_{\rm eff}$ \citep[cf.][]{KjeldsenBedding95,KjeldsenBedding11}, and on the stellar activity levels \citep[because elevated levels of activity affect detectability of the oscillations; see][]{starspotproxy,ChaplinActivity}. Here, we assumed solar values of temperature and activity to get the plotted solid-line trend. A recent account of alternative observational scaling relations aimed at predicting oscillation amplitudes is given in \citet{CorAscaling}. The frequency at which each $H_{\rm max,md}$ curve crosses $H_{\rm max,pred}$ corresponds to the seismic frequency $\nu_{\rm max,md}$ required for marginal detection of solar-like oscillations, given the observed noise background. The uncertainty in $\nu_{\rm max,md}$ is then defined by the intersection of $H_{\rm max,md}$ with the $\pm 1\sigma$ envelope associated with $H_{\rm max,pred}$. From Fig.~\ref{fig:exdetect}, we see that $\nu_{\rm max,md}$ decreases with increasing $m_{\rm Kep}$ as a result of the rising noise level.

The strength of the oscillations at the peak of the envelope is more commonly expressed as an equivalent amplitude, $A_{\rm max}$, for radial (i.e., $l\!=\!0$) modes rather than as a height, $H_{\rm max}$. We may convert the heights to amplitudes using \citep[][]{Kjeldsen08,BallotVis}:
\begin{equation}
A_{\rm max} = \sqrt{H_{\rm max}\,\Delta\nu/\xi} \, ,
\label{eq:httoamp}
\end{equation}
where $\xi$ corresponds to the total mode power per $\Delta\nu$ in units of $l\!=\!0$ power (viz., it measures the effective number of modes per order). The right-hand panel of Fig.~\ref{fig:exdetect} plots threshold amplitudes for detection, $A_{\rm max,md}$, and the expected trend in amplitude, $A_{\rm max,pred}$ (same line styles as in the left-hand panel). Unlike the threshold heights, the threshold amplitudes are seen to rise with increasing frequency. This is due to the presence of $\Delta\nu$ in Eq.~(\ref{eq:httoamp}), which increases with increasing $\nu_{\rm max}$ \citep[e.g.,][]{StelloScale}, more than offsetting the impact of a background that decreases with increasing frequency.

\subsection{Limits on surface gravities from marginal detection $\nu_{\rm max,md}$}\label{sec:limprop}
We may convert the marginal detection $\nu_{\rm max,md}$ into an equivalent marginal detection surface gravity, $g_{\rm md}$, for a given data set. The frequency of maximum amplitude, $\nu_{\rm max}$, is found to scale to very good approximation with the acoustic cutoff frequency \citep[][]{Brown91,KjeldsenBedding95,Belkacem11}, which, assuming an isothermal stellar atmosphere, gives a scaling relation for $\nu_{\rm max}$ in terms of surface gravity $g$ and effective temperature $T_{\rm eff}$. Solving for $g$ and normalizing by solar properties and parameters, one has
\begin{equation}
g \simeq {\rm g_{\sun}} \left(\frac{\nu_{\rm max}}{\nu_{\rm max,\sun}}\right) \left(\frac{T_{\rm eff}}{{\rm T_{eff,\sun}}}\right)^{1/2} \, ,
\label{eq:gscaling}
\end{equation}
with ${\rm g_{\sun}}\!=\!27402\:{\rm cm\,s^{-2}}$, $\nu_{\rm max,\sun}\!=\!3150\:{\rm \mu Hz}$ and $\rm{T_{eff,\sun}}\!=\!5777\:{\rm K}$. Hence, with independent knowledge of $T_{\rm eff}$, and substituting $\nu_{\rm max,md}$ for $\nu_{\rm max}$ in Eq.~(\ref{eq:gscaling}), we may estimate an equivalent marginal detection surface gravity, $g_{\rm md}$. It becomes apparent that the surface gravity $g$ is mainly dependent on $\nu_{\rm max}$, with the latter often taken as an indicator of the evolutionary state of a star. We should note that this so-called ``direct method'' of estimating stellar properties may lead to unphysically large uncertainties in the derived quantities since scaling relations are not constrained by the equations governing stellar structure and evolution. Alternatively, by comparing theoretical seismic quantities with the observed ones over a large grid of stellar models, very precise determinations of $\log g$ ($<\!0.05\:{\rm dex}$) can be obtained for F, G and K dwarfs \citep[e.g.,][]{CreeveyGaia}.

The accuracy of Eq.~(\ref{eq:gscaling}) has been the subject of several studies. For instance, \citet{HuberInterferometry} found no systematic deviations as a function of evolutionary state when testing the $\nu_{\rm max}$ scaling relation for a small sample of stars with available interferometric data. Based on a comparison involving about forty well-studied late-type pulsating stars with gravities derived using classical methods, \citet{MorelMiglio} found overall agreement with mean differences not exceeding $0.05\:{\rm dex}$. \citet{CreeveyGaia} studied sources of systematic errors in the determination of $\log g$ using grid-based methods and found possible biases of the order of $0.04\:{\rm dex}$. Since we are interested in computing an equivalent marginal detection value rather than the value itself, the quoted accuracies will not undermine the purpose of our study. We will adopt a conservative figure of $0.04\:{\rm dex}$ for the accuracy and adding it in quadrature to the uncertainties produced by Eq.~(\ref{eq:gscaling}).

We may, of course, estimate $\nu_{\rm max,md}$ for a {\it Kepler} data set irrespective of whether or not we have detected oscillations. For a star with data showing detected oscillations it must be the case that the observed maximum amplitude will be greater than or equal to the marginal amplitude for detection (allowing for uncertainty in the measurement), i.e., $A_{\rm max} \geqslant A_{\rm max,md}$. It must then also be the case that $\nu_{\rm max} \leqslant \nu_{\rm max,md}$. Inspection of Fig.~\ref{fig:exdetect} tells us that this is the case for 1-month-long observations of a solar twin made at $m_{\rm Kep}\!=\!10$, where we have assumed the solar value of $\nu_{\rm max}$, i.e., $\nu_{\rm max,\sun}$. Therefore, we now have an upper-limit estimate of a seismic property of the star. By using Eq.~(\ref{eq:gscaling}), we can translate this into an upper-limit estimate of a fundamental stellar property, to be specific, the surface gravity. In this case the direct determination of $\nu_{\rm max}$ from the observed oscillation power provides an estimate of $g$. However, in marginal cases the present analysis may serve as a consistency check of the reality of the detected oscillations.

Things get unquestionably more interesting when we consider data on stars for which we have failed to make a detection. Here, it must then be the case that $A_{\rm max,pred} < A_{\rm max,md}$ or, equivalently, $\nu_{\rm max} > \nu_{\rm max,md}$, meaning that the marginal detection $\nu_{\rm max,md}$ gives a lower-limit for the actual $\nu_{\rm max}$. Again, inspection of Fig.~\ref{fig:exdetect} suggests that we would fail to detect oscillations in solar twins with $m_{\rm Kep}\!=\!11$ and $m_{\rm Kep}\!=\!12$. Finally, use of Eq.~(\ref{eq:gscaling}) allows us to translate this lower-limit estimate of $\nu_{\rm max}$ into a lower-limit estimate of the surface gravity.

\subsection{Calibration of mode amplitude prediction $A_{\rm max,pred}$}\label{sec:calibration}
To establish a calibration for $A_{\rm max,pred}$, we used results on solar-type stars that have been observed in short cadence for at least three consecutive months from Q5 onwards as part of the KASC. We call this cohort 1. Stars in cohort 1 have moderate-to-high ${\rm S/N}$ in the p modes, which was one of the prerequisites for their selection for long-term observations by {\it Kepler}. The presence of solar-like oscillations in these stars had previously been confirmed based on observations made during the mission's survey phase \citep[e.g.,][]{ChaplinSci}. Light curves for these stars were prepared in the manner described by \citet{LightCurves} and were then high-pass filtered -- using a 1-day-cutoff triangular filter -- to remove any low-frequency power due to stellar activity and instrumental variability. We also analysed a smaller cohort of {\it Kepler} solar-type planet-candidate host stars with detected oscillations \citep[cf.][]{HuberEnsembleKOI}. We call this cohort 2, noting that cohorts 1 and 2 do not overlap. Preparation of the light curves for the stars in this second cohort differed from the procedure described above, although with no discernable impact on the homogeneity of the ensuing analysis. These light curves came from {\it Kepler} short-cadence data up to Q11. Specifically, transits needed to be corrected, since the sharp features in the time domain would cause significant power leakage from low frequencies into the oscillation spectrum in the frequency domain. This was achieved using a median filter with a length chosen according to the measured duration of the transit.

We extracted the global asteroseismic parameters $A_{\rm max}$ and $\nu_{\rm max}$ for the stars in both cohorts using the SYD \citep{HuberPipeline}, AAU \citep{CampantePipeline} and OCT \citep{HekkerPipeline} automated pipelines. Note that these pipelines were part of a thorough comparison exercise of complementary analysis methods used to extract global asteroseismic parameters of solar-type stars \citep{VernerComparison}. As a preliminary step to the calibration process per se, a validation of the extracted global asteroseismic parameters was carried out based on the prescription of \citet{VernerComparison}, which involved the rejection of outliers and a correction to the formal uncertainties returned by each of the analysis methods. The use of three pipelines in the validation of the extracted parameters was deemed sufficient, given the high ${\rm S/N}$ of the calibration stars. Firstly, we required that for each parameter determined for each star in either cohort, the results from at least two pipelines were contained within a range of fixed relative size centered on the median value (to be specific, $\pm21.5\,\%$ for $A_{\rm max}$ and $\pm10.5\,\%$ for $\nu_{\rm max}$). Results outside this range were iteratively removed until either all results were in agreement or fewer than two results remained. For each analysis method, only those stars with validated results for both parameters were retained. In addition, we demanded that the measured $\nu_{\rm max}\!>\!350\:{\rm \mu Hz}$, which approximately corresponds to the base of the red-giant branch \citep[e.g.,][]{HuberScaling}. Secondly, parameter uncertainties were recalculated by adding in quadrature the formal uncertainty and the standard deviation of the validated results over the contributing pipelines. The final (relative) median uncertainties in $A_{\rm max}$ and $\nu_{\rm max}$ for the three analysis methods are seen to lie within $6.9\text{--}8.0\,\%$ and $2.4\text{--}3.1\,\%$, respectively.

We used a Bayesian approach to calibrate a scaling relation for $A_{\rm max,pred}$ \citep[for more details, see][]{CorAscaling}. Two competing scaling relations (or models) were tested. Model $\mathcal{M}_1$ is based solely on the independent observables $\nu_{\rm max}$ and $T_{\rm eff}$:
\begin{equation}
\frac{A_{\rm max,pred}}{A_{\rm max,\sun}}\bigg|_{\mathcal{M}_1}=\beta \left(\frac{\nu_{\rm max}}{\nu_{\rm max,\sun}}\right)^{-s} 
	\left(\frac{T_{\rm eff}}{{\rm T_{eff,\sun}}}\right)^{3.5s-r} \, ,
\label{eq:Ascaling1}
\end{equation} 
where the solar maximum mode amplitude for the {\it Kepler} bandpass takes the value $A_{\rm max,\sun}\!=\!2.5\:{\rm ppm}$. The presence of the factor $\beta$ means that the model needs not to pass through the solar point. Model $\mathcal{M}_1$ has the same functional form as model $\mathcal{M}_{1,\beta}$ of \citet{CorAscaling}. The effective temperatures used here and in the remainder of this work were derived by \citet[][]{Pinsonneault}, who performed a recalibration of the KIC photometry in the Sloan Digital Sky Survey (SDSS) $griz$ filters using YREC models. Model $\mathcal{M}_2$, on the other hand, includes an extra exponential relation in the magnetic activity proxy, $\zeta_{\rm act}$:
\begin{equation}
\frac{A_{\rm max,pred}}{A_{\rm max,\sun}}\bigg|_{\mathcal{M}_2}=\beta \left(\frac{\nu_{\rm max}}{\nu_{\rm max,\sun}}\right)^{-s} 
	\left(\frac{T_{\rm eff}}{{\rm T_{eff,\sun}}}\right)^{3.5s-r} 
	{\rm e}^{m\,\zeta_{\rm act}} \, .
\label{eq:Ascaling2}
\end{equation} 
The magnetic activity proxy is described in Appendix \ref{sec:actproxy}. Four free parameters at most enter the Bayesian inference problem. We have adopted uniform priors for the model parameters $s$, $r$, and $m$, and a Jeffreys' prior for $\beta$, which results in an uniform prior for $\ln\beta$. Table \ref{tb:Ascaling1} lists the prior ranges adopted for each model parameter. Furthermore, we do not consider error-free independent variables, which means their relative uncertainties are properly taken into account by the likelihood function.

To proceed with the calibration of a scaling relation for $A_{\rm max,pred}$, we now look for an individual set of global asteroseismic parameters $A_{\rm max}$ and $\nu_{\rm max}$ (i.e., tracing back to a single analysis method), as opposed to some sort of average set, meaning that the parameters used in the calibration are fully reproducible. Furthermore, the fact that the extracted global asteroseismic parameters have been validated gives us confidence to use the output of any of the analysis methods in the calibration process. We opted for using the results arising from the SYD pipeline. The reasons behind this choice are simple: this pipeline generated the largest number of validated stars (163, of which 133 belong to cohort 1) and the broadest coverage in terms of the independent observables in Eqs.~(\ref{eq:Ascaling1}) and (\ref{eq:Ascaling2}): $350\!\lesssim\!\nu_{\rm max}\!\lesssim\!4400\:{\rm \mu Hz}$, $4950\!\lesssim\!T_{\rm eff}\!\lesssim\!7200\:{\rm K}$ and $10\!\lesssim\!\zeta_{\rm act}\!\lesssim\!1000\:{\rm ppm}$. This sample is dominated by main-sequence stars and so no attempt was made to derive separate scaling relations for $A_{\rm max,pred}$ based on the evolutionary state of the stars. In fact, for the ranges in $\nu_{\rm max}$ and $T_{\rm eff}$ being considered, the  observed logarithmic amplitudes are seen to vary approximately linearly with the logarithmic values of both $\nu_{\rm max}$ and $T_{\rm eff}$ -- as reproduced by models $\mathcal{M}_1$ and $\mathcal{M}_2$ -- with no abrupt change in slope, meaning that evolutionary influences are negligible \citep[see also][]{CorAscaling}. 

\begin{figure}
\epsscale{1.0}
\plotone{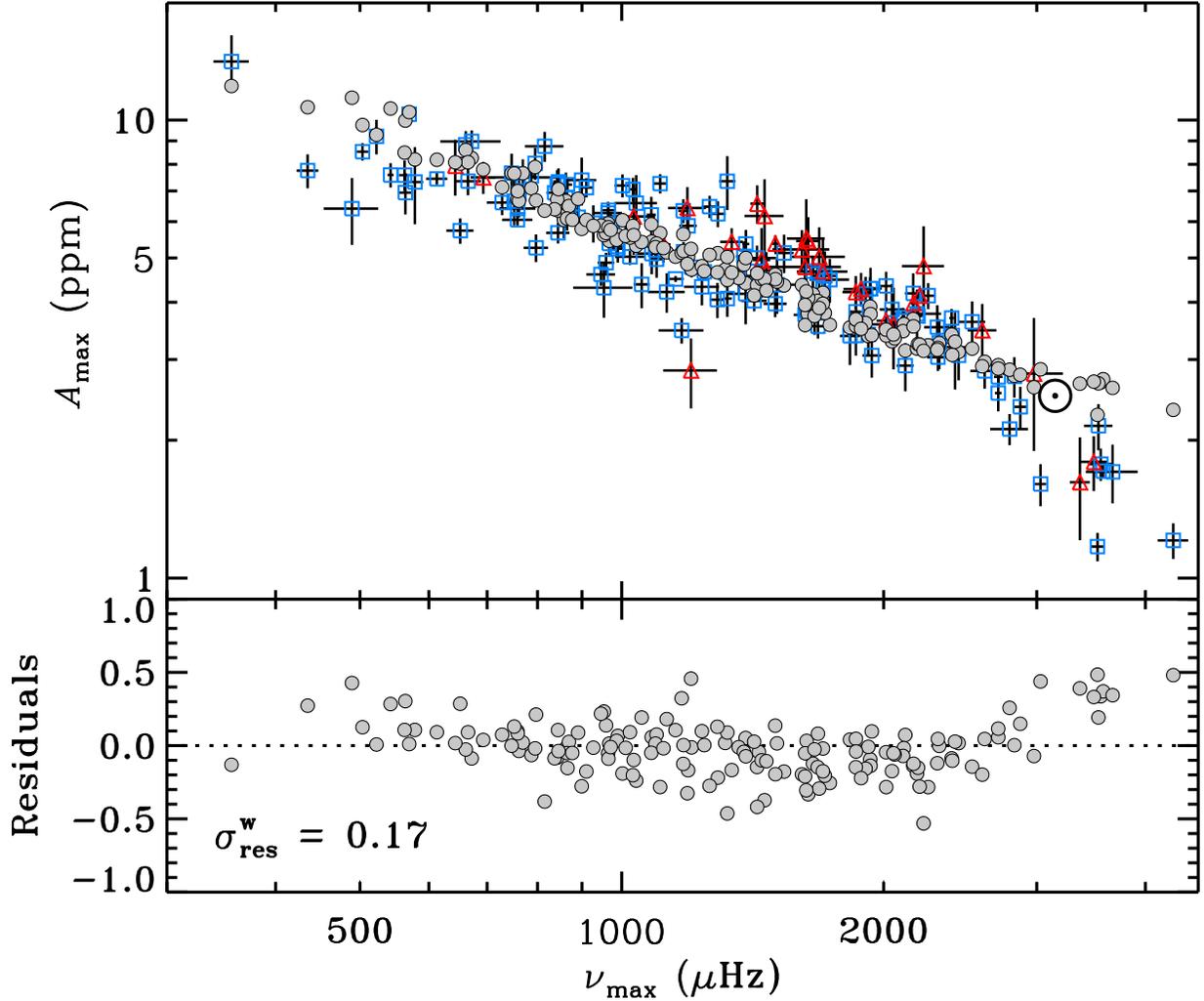}
\caption{\small Calibration of the amplitude scaling relation (or model) $\mathcal{M}_1$. Top panel: Predicted amplitudes (filled gray circles) are plotted against the measured $\nu_{\rm max}$. Observed amplitudes are shown in the background for stars both in cohort 1 (open blue squares) and cohort 2 (open red triangles). The solar symbol is placed according to the adopted solar reference values. Bottom panel: Relative residuals in the sense $({\rm Predicted}-{\rm Observed})/{\rm Predicted}$. Also shown is the weighted RMS of the relative residuals, an indicator of the quality of the fit.\label{fig:Ascaling1}}
\end{figure}

\begin{figure}
\epsscale{1.0}
\plotone{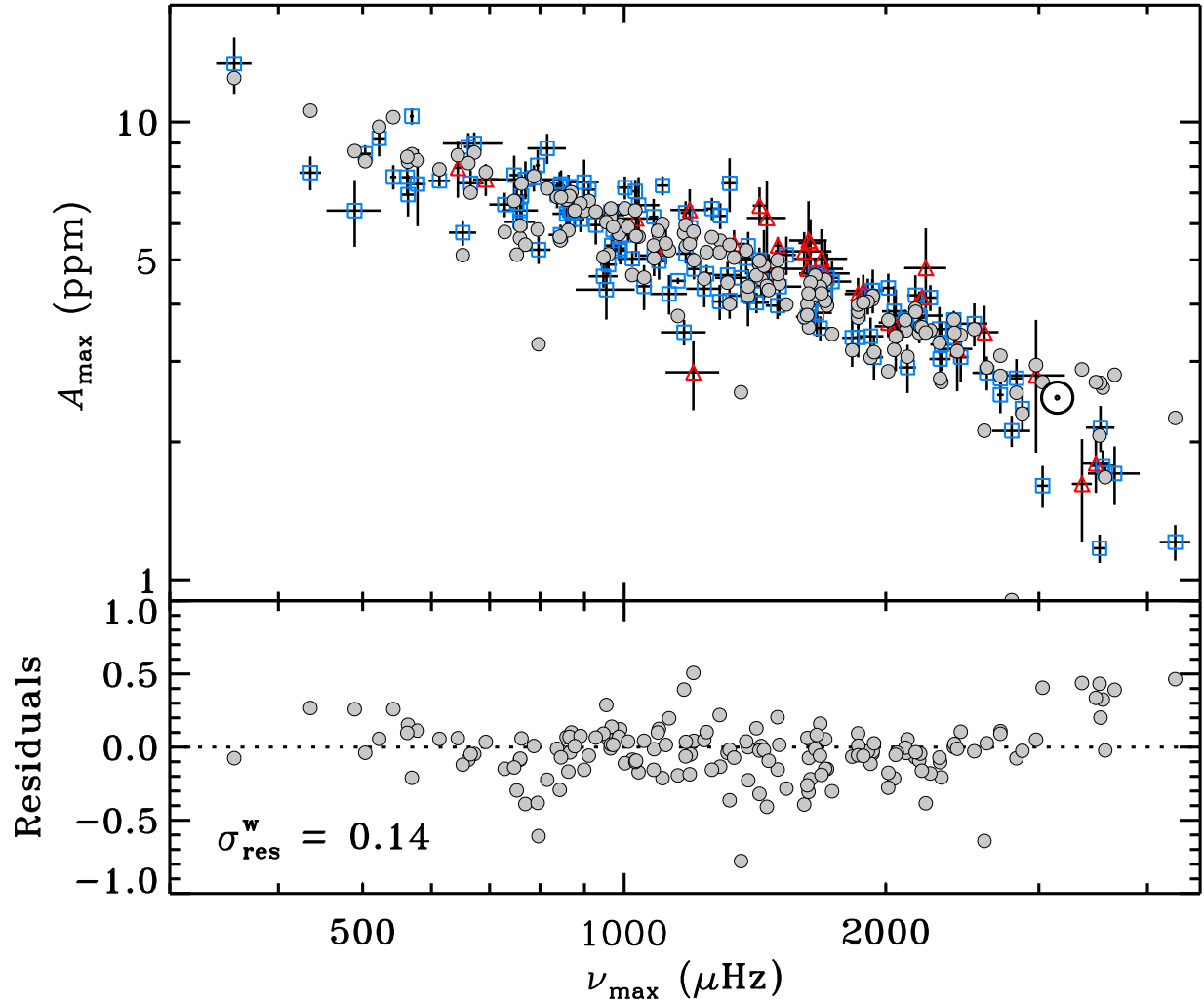}
\caption{\small Same as Fig.~\ref{fig:Ascaling1}, but for scaling relation (or model) $\mathcal{M}_2$.\label{fig:Ascaling2}}
\end{figure}

Tables \ref{tb:Ascaling2} and \ref{tb:Ascaling3} summarize the outcome of the Bayesian estimation of the model parameters. Note that the mean bias is much smaller than the observed scatter (i.e., $\bar{x}_{\rm res}\!\ll\!\sigma_{\rm res}^{\rm w}$). Figures \ref{fig:Ascaling1} and \ref{fig:Ascaling2} display the predicted and observed amplitudes, as well as the resulting relative residuals, for models $\mathcal{M}_1$ and $\mathcal{M}_2$, respectively. We note that the observed amplitudes of stars in cohort 2 are systematically higher than those in cohort 1. This is particularly noticeable at $\nu_{\rm max}\!\sim\!1500\:{\rm \mu Hz}$ in the top panels of Figs.~\ref{fig:Ascaling1} and \ref{fig:Ascaling2}. Stars in cohort 2 (having a magnitude distribution that peaks at $m_{\rm Kep}\!\simeq\!12$) are known to be globally fainter than members of cohort 1 (for which the magnitude distribution peaks at $m_{\rm Kep}\!\simeq\!11$), meaning that the high-frequency noise in the power spectrum will be greater. This leads to a selection bias in the measured amplitudes, viz., being fainter only those stars with the highest amplitudes will have detectable oscillations. In fact, cohort 2 consists primarily of slightly evolved F- and G-type stars \citep{HuberEnsembleKOI}, which have larger oscillation amplitudes than their unevolved counterparts.

Finally, we computed the so-called Bayes' factor in order to perform a formal statistical comparison of the two competing models. Computing the Bayes' factor in favor of model $\mathcal{M}_2$ over model $\mathcal{M}_1$ (i.e., $B_{21}\!\equiv\!\mathcal{E}_{\mathcal{M}_2}/\mathcal{E}_{\mathcal{M}_1}$, the ratio of the Bayesian evidences) gave a logarithmic factor of $\ln B_{21}\!\gg\!1$, decisively favoring model $\mathcal{M}_2$ over model $\mathcal{M}_1$ \citep[][]{Jeffreys61}. Note that we are not saying that model $\mathcal{M}_2$ is physically more meaningful, but rather statistically more likely. This in turn renders statistical significance to the inclusion of an extra dependence on $\zeta_{\rm act}$ in model $\mathcal{M}_2$. The negative value of model parameter $m$ implies that amplitudes of solar-like oscillations are suppressed in stars with increased levels of surface magnetic activity. This corroborates the conclusions of \citet{ChaplinActivity} -- where an exponential relation was also considered -- and strengthens the quantitative results on stellar activity and amplitudes presented by \citet{HuberScaling}. Henceforth, we use the calibration for $A_{\rm max,pred}$ based on model $\mathcal{M}_2$ (Eq.~\ref{eq:Ascaling2}).


\begin{table}[h]
\begin{center}
\caption{\small Prior ranges adopted in the Bayesian estimation of the model parameters.\label{tb:Ascaling1}}
\begin{tabular}{lr}
\tableline\tableline
Parameter & Prior range \\
\tableline
$\ln\beta$ & $\left[-0.5,0.5\right]$ \\
$s$ & $\left[0.5,1.0\right]$ \\
$r$ & $\left[2.4,4.4\right]$ \\
$m$ & $\left[-0.2,0.2\right]$ \\
\tableline
\end{tabular}
\end{center}
\end{table}


\begin{table}[h]
\begin{center}
\caption{\small Expected values of the inferred model parameters and their associated $68.3\,\%$ Bayesian credible regions. The logarithm of the Bayesian evidence, $\ln\mathcal{E}$, the mean relative residuals, $\bar{x}_{\rm res}$, and the weighted RMS of the relative residuals, $\sigma_{\rm res}^{\rm w}$, are also reported.\label{tb:Ascaling2}}
\begin{tabular}{lccccccc}
\tableline\tableline
Model & $\ln\beta$ & $s$ & $r$ & $m$ & $\ln\mathcal{E}$ & $\bar{x}_{\rm res}$ & $\sigma_{\rm res}^{\rm w}$ \\
&&&& ($\times10^{-4}\:{\rm ppm^{-1}}$) &&& \\
\tableline
$\mathcal{M}_1$ & $0.09^{+0.01}_{-0.02}$ & $0.71^{+0.01}_{-0.01}$ & $3.42^{+0.11}_{-0.10}$ & $\cdots$ & $-469.8$ & $-0.015\pm0.001$ & $0.17$ \\
$\mathcal{M}_2$ & $0.22^{+0.02}_{-0.02}$ & $0.68^{+0.02}_{-0.02}$ & $2.83^{+0.14}_{-0.13}$ & $-9.5^{+0.9}_{-0.8}$ & $-24.3$ & $-0.038\pm0.001$ & $0.14$ \\
\tableline
\end{tabular}
\end{center}
\end{table}


\begin{table}[h]
\begin{center}
\caption{\small Correlation coefficients for each pair of model parameters.\label{tb:Ascaling3}}
\begin{tabular}{lcccccc}
\tableline\tableline
Model & $s$ vs.~$r$ & $s$ vs.~$\ln\beta$ & $s$ vs.~$m$ & $r$ vs.~$\ln\beta$ & $r$ vs.~$m$ & $\ln\beta$ vs.~$m$ \\
\tableline
$\mathcal{M}_1$ & $0.16$ & $-0.81$ & $\cdots$ & $0.31$ & $\cdots$ & $\cdots$ \\
$\mathcal{M}_2$ & $0.40$ & $-0.75$ & $0.01$ & $0.03$ & $0.15$ & $-0.39$ \\
\tableline
\end{tabular}
\end{center}
\end{table}


\section{Results}
We have applied the method described above to three non-overlapping cohorts of stars, namely, to solar-type stars with detected oscillations that were observed as part of the KASC (cohort 1, as before), and to solar-type planet-candidate host stars both with (cohort 2, as before) and without (cohort 3) detected oscillations. For every data set in each of these three cohorts, we computed the mode amplitude threshold $A_{\rm max,md}$ as a function of frequency (cf.~Appendix \ref{sec:threshold}). We compared this to the mode amplitude prediction $A_{\rm max,pred}$ for that star (computed using the known $T_{\rm eff}$ and a proxy measure of its activity; cf.~Sect.~\ref{sec:calibration}), which yielded the sought-for $\nu_{\rm max,md}$ (cf.~Sect.~\ref{sec:overview}) and $g_{\rm md}$ (cf.~Sect.~\ref{sec:limprop}).

\subsection{Solar-type stars with detected oscillations}\label{sec:resosc}
\begin{figure}
\centering
\includegraphics*[scale=0.55]{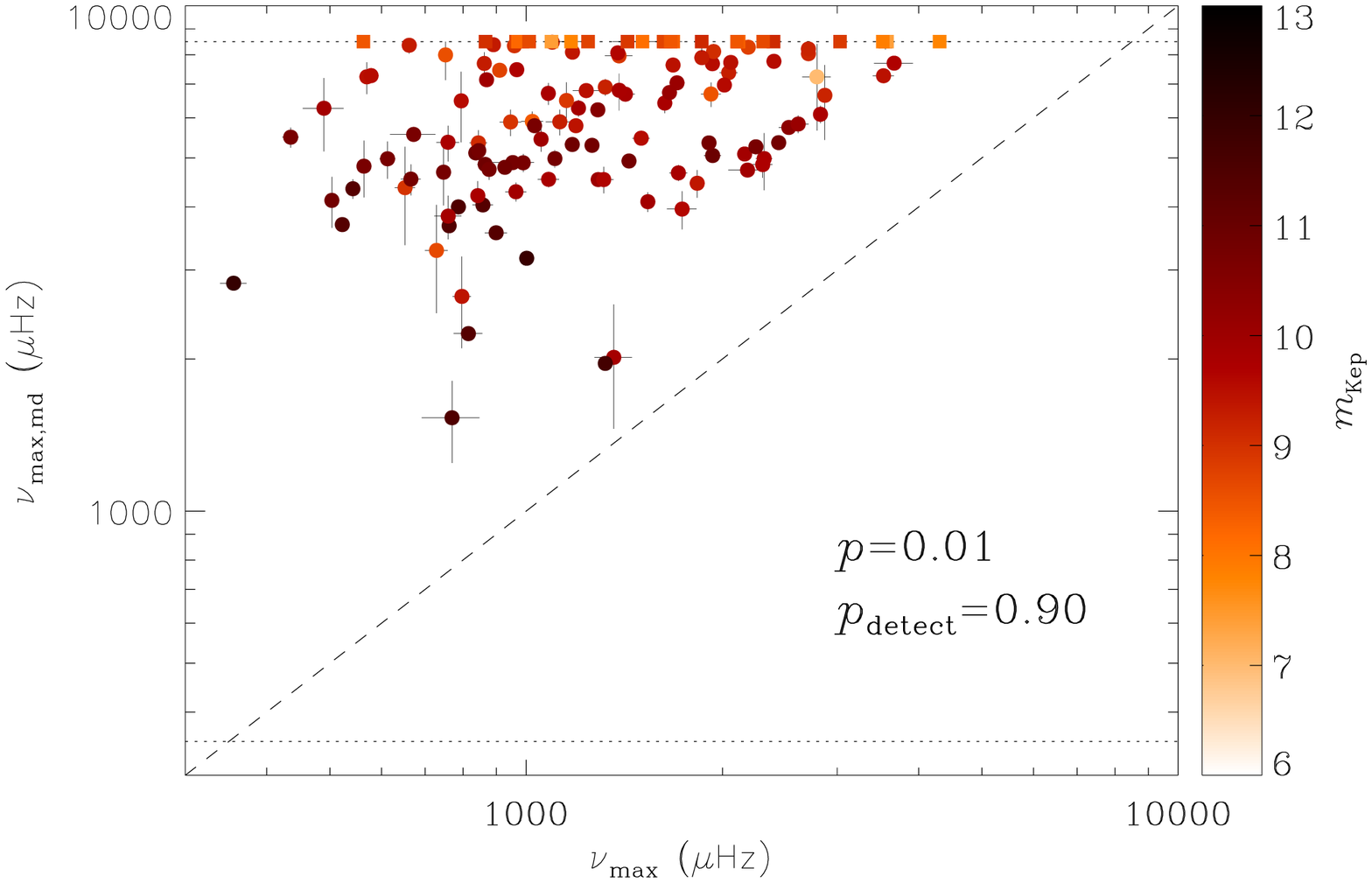}
\includegraphics*[scale=0.55]{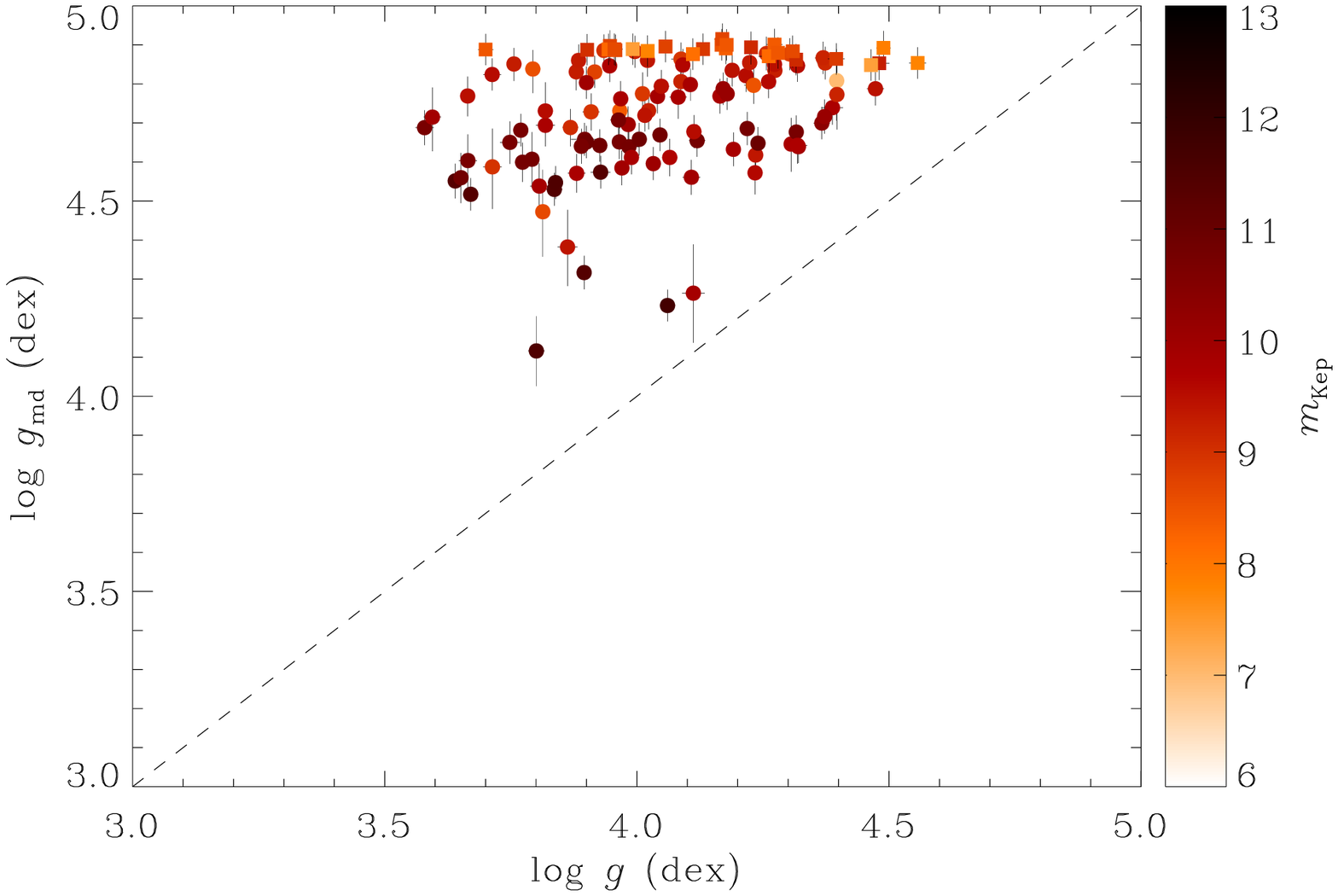}
\caption{\small Top panel: Computed $\nu_{\rm max,md}$ versus the measured $\nu_{\rm max}$ (from the SYD pipeline) for solar-type stars showing oscillations observed as part of the KASC (cohort 1). Horizontal dotted lines delimit the range in frequency that has been tested in the determination of $\nu_{\rm max,md}$ (to be specific, from $350\:{\rm \mu Hz}$ to the Nyquist frequency for {\it Kepler} short-cadence data, $\nu_{\rm Nyq}\!\sim\!8496\:{\rm \mu Hz}$). The adopted $p$-value and detection probability are indicated. Bottom panel: Corresponding marginal detection gravities, $g_{\rm md}$, versus the seismically determined gravities from \citet{ChaplinFundProp}. In both panels, the dashed line represents the one-to-one relation. Filled squares indicate that the determination of $\nu_{\rm max,md}$ has saturated (viz., $\nu_{\rm max,md}$ equals the upper boundary of the tested frequency range). Points are colored according to magnitude.\label{fig:wg1osc}}
\end{figure}

As a sanity check on the marginal detection methodology, we began by analyzing solar-type stars for which the presence of oscillations had previously been confirmed. The top panel of Fig.~\ref{fig:wg1osc} shows the computed $\nu_{\rm max,md}$ versus the observed $\nu_{\rm max}$ for stars in cohort 1. We depict only stars with validated values of $\nu_{\rm max}$ coming from the SYD pipeline. Since these stars have detections, our sanity check amounts to verifying that $\nu_{\rm max} \leqslant \nu_{\rm max,md}$, which is indeed found to be the case (with all points lying well above the one-to-one line). Notice that, as a general rule, the brighter the star the higher is $\nu_{\rm max,md}$ (and the farther it lies above the one-to-one line), thus making it possible to detect oscillations with even the highest $\nu_{\rm max}$. The bottom panel of Fig.~\ref{fig:wg1osc} plots the corresponding marginal detection surface gravities, $g_{\rm md}$, versus the seismically determined gravities from \citet{ChaplinFundProp}. It must be the case for these stars that $g \leqslant g_{\rm md}$ (cf.~Eq.~\ref{eq:gscaling}). Again, the sanity check is seen to hold well.

\begin{figure}
\centering
\includegraphics*[scale=0.55]{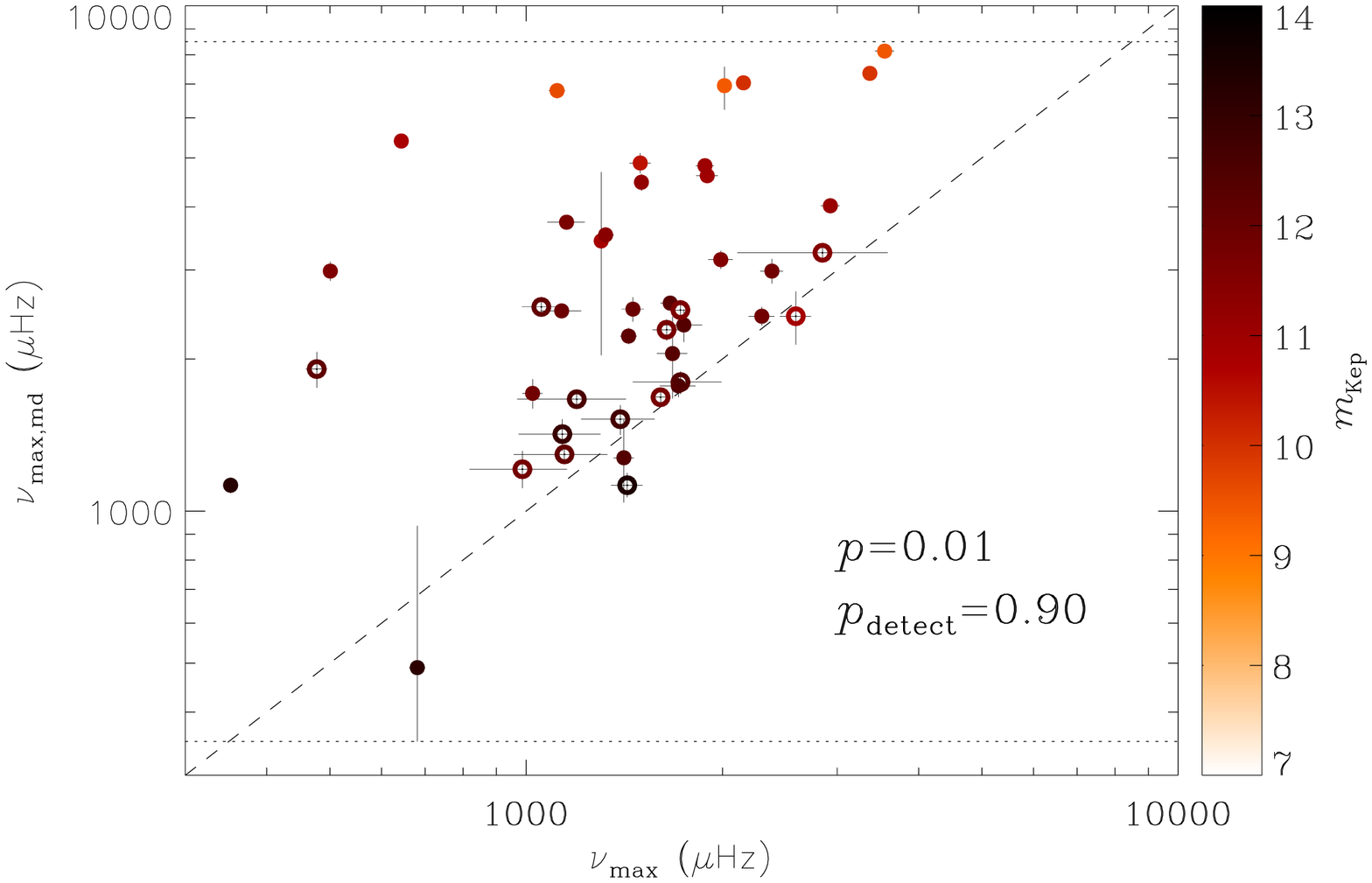}
\includegraphics*[scale=0.55]{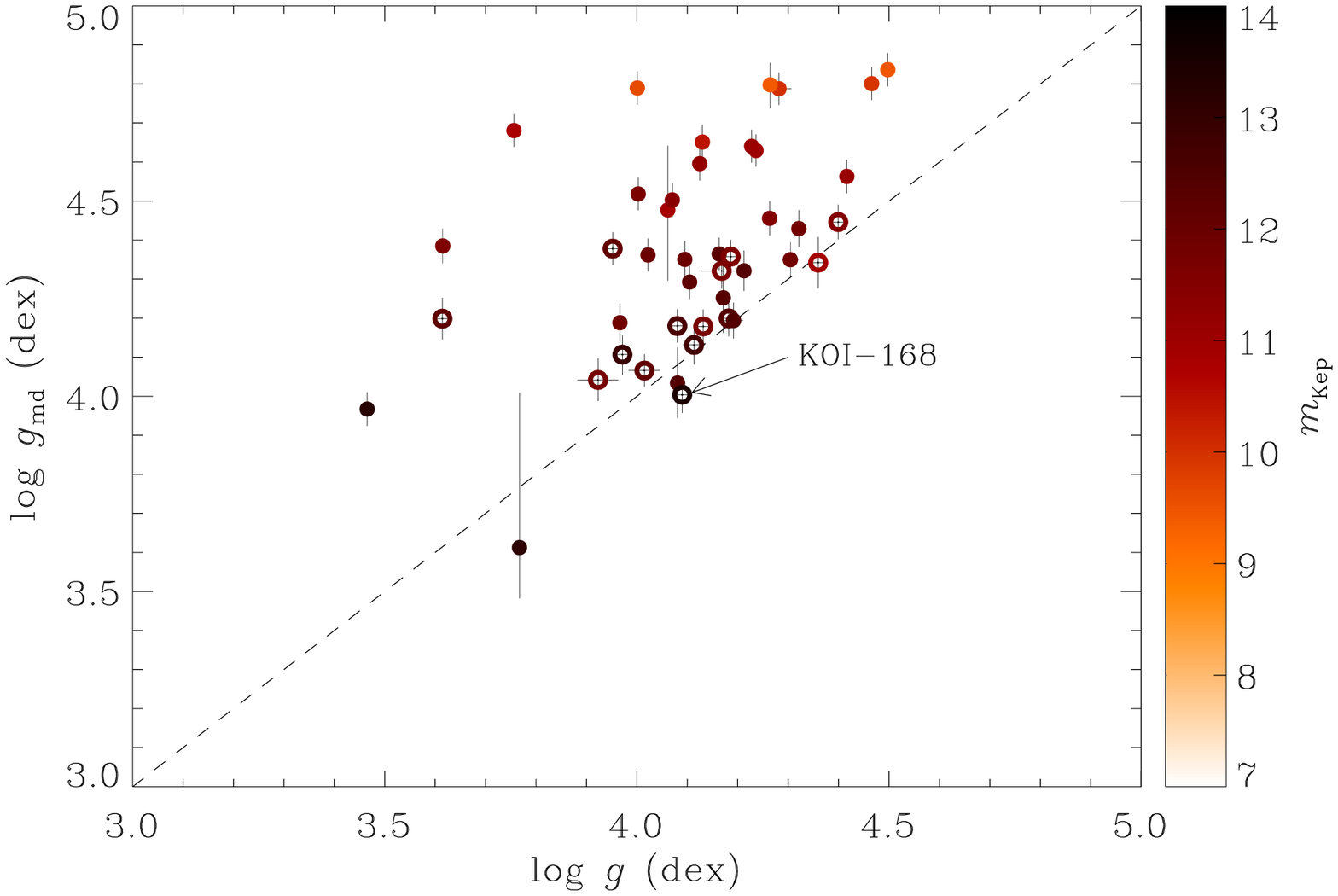}
\caption{\small Top panel: Computed $\nu_{\rm max,md}$ versus the measured $\nu_{\rm max}$ \citep[from][]{HuberEnsembleKOI} for solar-type planet-candidate host stars showing oscillations (cohort 2). Horizontal dotted lines delimit the range in frequency that has been tested in the determination of $\nu_{\rm max,md}$ (to be specific, from $350\:{\rm \mu Hz}$ to the Nyquist frequency, $\nu_{\rm Nyq}$). The adopted $p$-value and detection probability are indicated. Bottom panel: Corresponding marginal detection gravities, $g_{\rm md}$, versus the seismically determined gravities from \citet{HuberEnsembleKOI}. In both panels, the dashed line represents the one-to-one relation. Points are colored according to magnitude. Stars for which the estimation of $\nu_{\rm max}$ was deemed unreliable by those authors are represented by open symbols.\label{fig:koiosc}}
\end{figure}

A similar check was done that focused on stars belonging to cohort 2 (see Fig.~\ref{fig:koiosc}). The observed $\nu_{\rm max}$ values in the top panel are taken from \citet{HuberEnsembleKOI} and limited to $\nu_{\rm max}\!>\!350\:{\rm \mu Hz}$. Note that for some of the stars, the estimation of $\nu_{\rm max}$ was deemed unreliable by those authors due to the low signal-to-noise ratio in the oscillation spectrum (represented by open symbols). Not all of the depicted stars have actually entered the calibration (as is the case in Fig.~\ref{fig:wg1osc}), and we have thus been careful to guarantee that they comply with the considered ranges in $T_{\rm eff}$ and $\zeta_{\rm act}$. The reference (seismically determined) surface gravities in the bottom panel of Fig.~\ref{fig:koiosc} also come from \citet{HuberEnsembleKOI}. Once more, the sanity check is seen to hold, while resulting in less conservative estimates of $\nu_{\rm max,md}$ and $g_{\rm md}$, that is to say, they lie closer to the one-to-one line. We attribute this mainly to the fact that stars in cohort 2 are globally fainter than stars in cohort 1 (as already mentioned in Sect.~\ref{sec:calibration}). The sole apparent outlier in the bottom panel of Fig.~\ref{fig:koiosc} (in the sense of not being consistent with the one-to-one relation at the $1\sigma$ level) corresponds to KOI-168 ($m_{\rm Kep}\!=\!13.44$), which has an unreliable estimation of $\nu_{\rm max}$.


\subsection{Solar-type stars without detected oscillations}\label{sec:resnonosc}
We now turn our attention to the analysis of {\it Kepler} solar-type planet-candidate host stars with no detected oscillations (cohort 3), with the intention of providing lower-limit $\log g$ estimates. The light curves for these targets come from available {\it Kepler} short-cadence data up to Q14. Preparation of these light curves has been done in the same way as for targets in cohort 2 (cf.~Sect.~\ref{sec:calibration}). Selected targets comply with the ranges in $T_{\rm eff}$ and $\zeta_{\rm act}$ adopted in the calibration process. This gives a total of 453 targets.

\begin{figure}
\epsscale{1.0}
\plotone{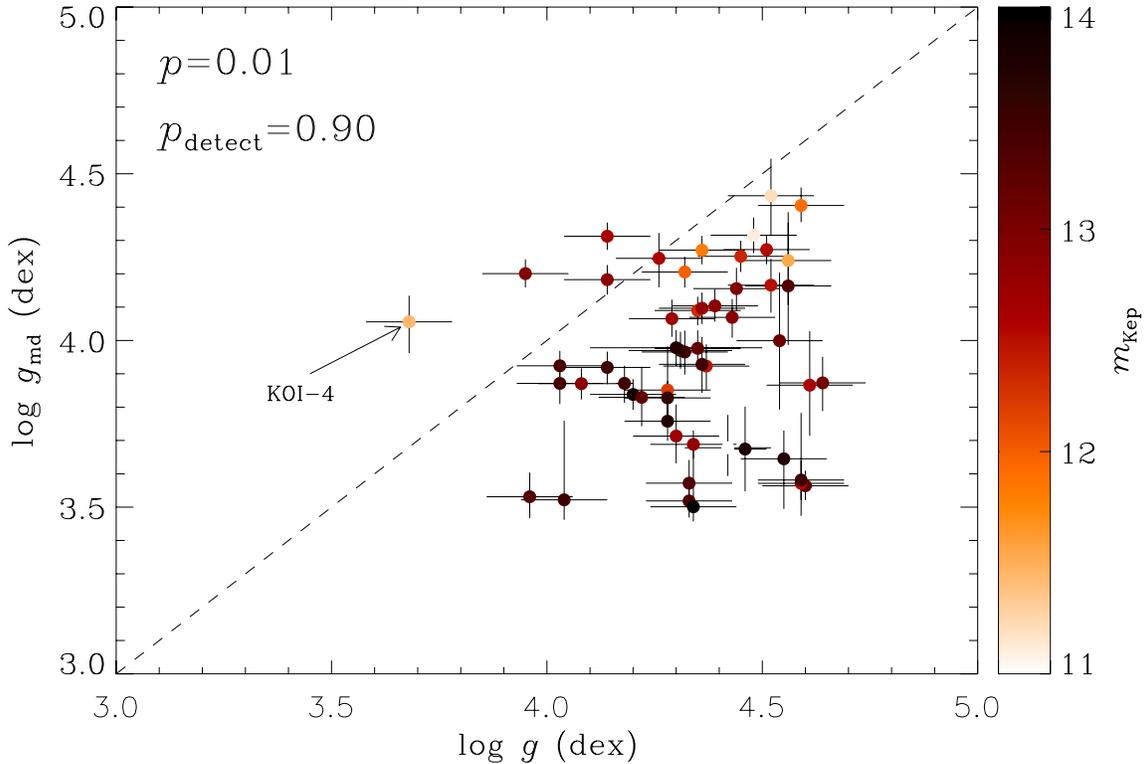}
\caption{\small Computed marginal detection surface gravities, $g_{\rm md}$, versus the spectroscopic values from \citet{Buchhave} for {\it Kepler} solar-type planet-candidate host stars with no detected oscillations. The dashed line represents the one-to-one relation. Points are colored according to magnitude. The adopted $p$-value and detection probability are indicated.\label{fig:koinonoscBuch}}
\end{figure}

We compared the marginal detection surface gravities, $g_{\rm md}$, with the spectroscopic values from \citet{Buchhave} for stars common to both sets (see Fig.~\ref{fig:koinonoscBuch}). Of the 69 common stars, 16 are saturated (viz., the underlying $\nu_{\rm max,md}$ equals the lower boundary of the tested frequency range, set at $350\:{\rm \mu Hz}$) and have not been plotted. This leaves 53 useful data points. Of these, 49 (50 if we allow for the uncertainty in $\log g$ alone) fall below the one-to-one line. This general trend is to be expected for stars with no detected oscillations, for which $g > g_{\rm md}$. A potential application of our method could be to identify those KOIs that were misclassified as subgiants by \citet{Buchhave}. Taking $\log g\!=\!3.85\:{\rm dex}$ as an indicative cutoff between main-sequence stars and subgiants, we point out that KOI-4 ($\log g\!=\!3.68\pm0.10\:{\rm dex}$, $\log g_{\rm md}\!=\!4.06^{+0.08}_{-0.09}\:{\rm dex}$) has possibly been misclassified as a subgiant. We should note that this cutoff is dependent on stellar mass and metallicity, as well as on the amount of overshooting when a convective core is present. Specifying this cutoff at $\log g\!=\!3.85\:{\rm dex}$ serves merely to illustrate this potential application of the method.

\begin{figure}
\epsscale{1.0}
\plotone{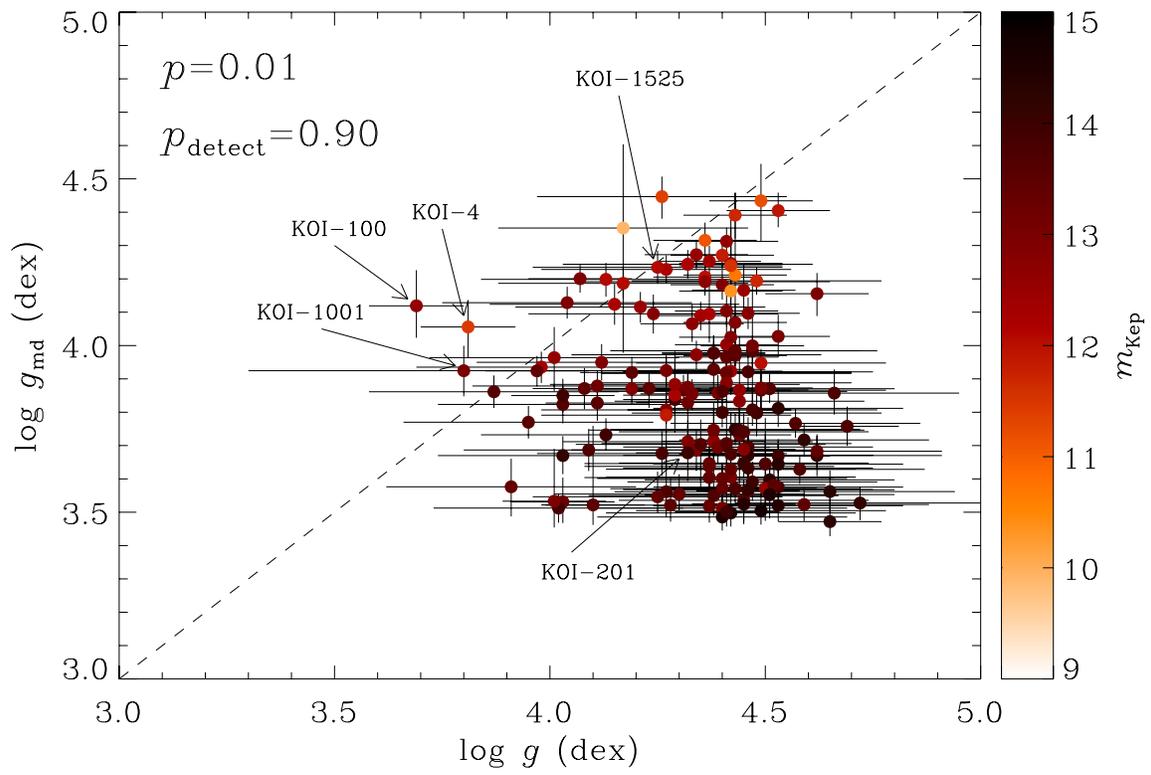}
\caption{\small Same as Fig.~\ref{fig:koinonoscBuch}, but for the comparison with the $\log g$ values from \citet{BatalhaCandidates3}.\label{fig:koinonoscBat}}
\end{figure}

Comparison with the $\log g$ values from \citet{BatalhaCandidates3} for stars common to both sets is shown in Fig.~\ref{fig:koinonoscBat}. These values do not come with an associated error bar and so we have adopted the standard deviations of the residuals in table 4 of \citet{HuberEnsembleKOI} as notional error bars. Accordingly, for $\log g\!<\!3.85\:(\log g\!>\!3.85)$: $\sigma_{\log g}\!=\!0.11\:(0.12)$ for stars with spectroscopic follow-up, otherwise $\sigma_{\log g}\!=\!0.50\:(0.29)$ for stars with available KIC parameters only\footnote{Note that an uncertainty of $0.4\:{\rm dex}$ is typically assumed in the KIC for $\log g$ \citep[e.g.,][]{VernerKIC}.}. Of the 354 stars common to both sets, 190 are saturated. This leaves 164 useful data points. Of these, $94\,\%$ of the points ($98\,\%$ if we allow for the uncertainty in $\log g$ alone) fall below the one-to-one line. Three KOIs have possibly been misclassified as subgiants by \citet{BatalhaCandidates3}: KOI-4 ($\log g\!=\!3.81\pm0.11\:{\rm dex}$, $\log g_{\rm md}\!=\!4.06^{+0.08}_{-0.09}\:{\rm dex}$), KOI-100 ($\log g\!=\!3.69\pm0.11\:{\rm dex}$, $\log g_{\rm md}\!=\!4.12^{+0.11}_{-0.10}\:{\rm dex}$) and KOI-1001 ($\log g\!=\!3.80\pm0.50\:{\rm dex}$, $\log g_{\rm md}\!=\!3.92^{+0.07}_{-0.08}\:{\rm dex}$). Notice that KOI-4 has once again been listed as a misclassified subgiant. 

We propose lower-limit $\log g$ estimates for {\it Kepler} solar-type planet-candidate host stars with no detected oscillations, as given in Table \ref{tb:finaltable}. We have discarded targets for which the determination of $\nu_{\rm max,md}$ has saturated, since the associated marginal detection surface gravities are not bona fide lower-limit $\log g$ estimates. This leaves 220 targets of the potential 453 targets mentioned above. The faintest target in this subset of 220 stars has $m_{\rm Kep}\!=\!14.73$, in contrast with $m_{\rm Kep}\!=\!16.42$ for the full cohort. Clearly, target saturation is closely linked to faint magnitudes (cf.~Fig.~\ref{fig:exdetect}). The final median uncertainty in $\log g_{\rm md}$ is $0.06\:{\rm dex}$ (or $0.04\:{\rm dex}$ if we do not include the figure of $0.04\:{\rm dex}$ added in quadrature to the uncertainties produced by Eq.~\ref{eq:gscaling}).

Inspection of Fig.~\ref{fig:koinonoscBat} reveals that, in most cases, the proposed $\log g_{\rm md}$ are significantly smaller than the $\log g$ values. This is particularly true for the faintest stars, as can be seen around $\log g\!\sim\!4.4\:{\rm dex}$. In order to evaluate the performance of our method, we computed $\Delta \log g$, i.e., the difference between the $\log g$ from \citet{BatalhaCandidates3} and the proposed $\log g_{\rm md}$ (individual $\Delta \log g$ are given in Table \ref{tb:finaltable}; mean $\Delta \log g$ and associated scatter are given in Table \ref{tb:Deviation}). We started by categorizing the stars into two uniform intervals in $\log g$. We restricted ourselves to main-sequence stars (i.e., $\log g\!\gtrsim\!3.85\:{\rm dex}$), since these make up the vast majority of the plotted data points. We further distinguished between bright ($m_{\rm Kep}\!\leq\!12.4$) and faint ($m_{\rm Kep}\!>\!12.4$) targets, with $m_{\rm Kep}\!=\!12.4$ being the magnitude of the faintest host star with detected oscillations among those close to the main sequence \citep{HuberEnsembleKOI}. The mean $\Delta \log g$ are to be compared with the notional uncertainties on $\log g$ quoted above (i.e., $\sigma_{\log g}\!=\!0.12$ for stars with spectroscopic follow-up, otherwise $\sigma_{\log g}\!=\!0.29$ for stars with available KIC parameters only). We first notice an increase in the mean $\Delta \log g$ with $\log g$. This is in part tied to the variation with $\log g$ of the maximum allowed excursion for $\Delta \log g$, imposed by the lower boundary of the tested frequency range (which translates into a $\log g_{\rm md}\!\sim\!3.5\:{\rm dex}$ floor). The effect of the stellar magnitude is conspicuous. The $\Delta \log g$ for bright stars ($15\,\%$ of the plotted data points in Fig.~\ref{fig:koinonoscBat}) are on average commensurate with or smaller than the magnitude of the quoted uncertainties on $\log g$, rendering the $\log g_{\rm md}$ estimates useful. On the other hand, the $\Delta \log g$ for faint stars are on average considerably larger than (for $4.30\!\leq\!\log g\!\leq\!4.75$; $62\,\%$ of the data points) or at most commensurate with (for $3.85\!\leq\!\log g\!<\!4.30$; $21\,\%$ of the data points) the uncertainties on $\log g$. Stellar magnitude then strongly affects the usefulness of the $\log g_{\rm md}$ estimates, especially for faint stars with $\log g\!\geq\!4.30$ in the \citet{BatalhaCandidates3} catalog (e.g., see locus of KOI-201 in Fig.~\ref{fig:koinonoscBat}). We have not yet mentioned the effect of the length of the observations, which may explain part of the observed scatter. This effect is, nonetheless, considerably weaker than that of the stellar magnitude. This is especially true for the faintest stars, for which multi-year observations are needed to produce a noticeable rise in the detection probability \citep[cf.][]{ChaplinDetect}. In such cases, an upgrade of a few observing months would produce no apparent change in the computed $\log g_{\rm md}$.


\begin{deluxetable}{lccccc}
\tablecolumns{6}
\tablewidth{0pt}
\tablenum{5}
\tablecaption{\small Mean differences between $\log g$ values from \citet{BatalhaCandidates3} and proposed $\log g_{\rm md}$.\label{tb:Deviation}}
\tablehead{
\colhead{} & \multicolumn{2}{c}{$3.85\!\leq\!\log g\!<\!4.30$} & \colhead{} & 
\multicolumn{2}{c}{$4.30\!\leq\!\log g\!\leq\!4.75$} \\
\cline{2-3} \cline{5-6} \\
\colhead{} & \colhead{$m_{\rm Kep}\!\leq\!12.4$} & \colhead{$m_{\rm Kep}\!>\!12.4$} & \colhead{} &
\colhead{$m_{\rm Kep}\!\leq\!12.4$} & \colhead{$m_{\rm Kep}\!>\!12.4$}}
\startdata
$\Delta \log g$ (dex) & $0.01\pm0.02\:(0.12)$ & $0.29\pm0.01\:(0.22)$ & & $0.19\pm0.02\:(0.15)$ & 
$0.658\pm0.006\:(0.25)$ \\
\enddata
\tablecomments{Error bars are given by the standard error of the mean. Numbers in brackets are the standard deviation of the residuals.}
\end{deluxetable}


\section{Summary and discussion}\label{sec:discussion}
We have presented a novel method for placing limits on the seismic (i.e., $\nu_{\rm max}$) and hence fundamental properties (i.e., $\log g$) of {\it Kepler} targets in the data on which we have not been able to detect signatures of solar-like oscillations. For a given noise background and length of the observations, we showed how to estimate the (frequency-dependent) maximum mode amplitude, $A_{\rm max,md}$, required to make a marginal detection against that background. We then established a calibration for the predicted maximum mode amplitude, $A_{\rm max,pred}$, as a function of $\nu_{\rm max}$, $T_{\rm eff}$ and $\zeta_{\rm act}$. Comparing $A_{\rm max,md}$ to $A_{\rm max,pred}$ then yielded $\nu_{\rm max,md}$. Finally, use of a scaling relation allowed $\log g_{\rm md}$ to be computed.

A proxy of the level of stellar activity, $\zeta_{\rm act}$, has been introduced that can be obtained directly from the light curve. This proxy could prove useful in future activity-cycle studies. An important byproduct of the calibration process for $A_{\rm max,pred}$ has been the confirmation that amplitudes of solar-like oscillations are suppressed in stars with increased levels of surface magnetic activity.

As a sanity check, the method was first tested on two distinct cohorts of stars showing detected oscillations, namely, on solar-type stars observed as part of the KASC and on solar-type KOIs. For a star with detected oscillations it must be the case that $\nu_{\rm max} \leqslant \nu_{\rm max,md}$ or, equivalently, $g \leqslant g_{\rm md}$. Therefore, the marginal detection $g_{\rm md}$ must be an upper-limit estimate of the actual surface gravity. This sanity check was seen to hold well.

Conversely, for a star for which we failed to make a detection, it must then be the case that $\nu_{\rm max} > \nu_{\rm max,md}$ or, equivalently, $g > g_{\rm md}$. Therefore, the marginal detection $g_{\rm md}$ must now be a lower-limit estimate of the actual surface gravity. While bearing in mind the existing limitations on the determination of accurate $\log g$ estimates for {\it Kepler} planet-candidate host stars (cf.~Sect.~\ref{sec:intro}), we nonetheless compared our marginal detection $\log g_{\rm md}$ with the $\log g$ values from \citet{Buchhave} and \citet{BatalhaCandidates3}, largely confirming our expectations of finding the condition $g > g_{\rm md}$ to be generally satisfied. We have proposed lower-limit $\log g$ estimates for 220 solar-type KOIs ($m_{\rm Kep}\!<\!15$) with no detected oscillations. We evaluated the performance of our method based on a comparison of the (mean) deviation, $\Delta \log g$, of the proposed $\log g_{\rm md}$ from the $\log g$ values in \citet{BatalhaCandidates3}, with the quoted uncertainties on $\log g$. As a result, we pointed out the reduced usefulness of $\log g_{\rm md}$ estimates for faint stars (i.e., $m_{\rm Kep}\!>\!12.4$) with $\log g\!\geq\!4.30$ in the \citet{BatalhaCandidates3} catalog, which comprise $62\,\%$ of the plotted data points in Fig.~\ref{fig:koinonoscBat}. We should, however, note the potential biases affecting stellar properties in that catalog. \citet{HuberEnsembleKOI} showed surface gravities for subgiant and giant host stars based on high-resolution spectroscopy to be systematically overestimated. Besides, surface gravities for unevolved stars based on KIC parameters were also found to be systematically overestimated in that catalog. A correction for these biases (not applied in this work) would bring the plotted data points in Fig.~\ref{fig:koinonoscBat} closer to the one-to-one line, thus improving the perceived performance of our method. Furthermore, we should note that a KIC $\log g\!>\!4$ does not necessarily mean that the possibility of a star being a giant is ruled out \citep{Mann12}. Consequently, for a typical star with $\log g\!\sim\!4.4$ and $\log g_{\rm md}\!\sim\!3.5$ (see Fig.~\ref{fig:koinonoscBat}), the proposed marginal detection surface gravity may still be a useful constraint by ruling out the giant scenario.  

The information contained in the $\log g_{\rm md}$ estimates is likely to be useful in the characterization of the corresponding candidate planetary systems, namely, by helping constrain possible false-positive scenarios (and thus promote candidates to genuine exoplanets), and/or by constraining the transit model for systems that have already been validated. We give two specific examples that illustrate the potential use of the proposed $\log g_{\rm md}$ estimates, one being characterized by $\Delta \log g\!\sim\!0$ (KOI-1525) and the other by $\Delta \log g\!<\!0$ (KOI-100):
\begin{enumerate}

\item KOI-1525 (see Fig.~\ref{fig:koinonoscBat}) has no spectroscopic follow-up and a $\log g$ based on the KIC: $\log g\!=\!4.2\pm0.4\:{\rm dex}$ (where we have assumed a typical KIC uncertainty of $0.4\:{\rm dex}$ for $\log g$). Hence, this star could either be a subgiant or a main-sequence star, with a radius uncertainty of about $60\,\%$. The non-detection of oscillations yields $\log g_{\rm md}\!=\!4.23^{+0.05}_{-0.05}\:{\rm dex}$, thus ruling out the subgiant scenario. This is likely a main-sequence star of spectral type F ($T_{\rm eff}\!=\!6905\pm87\:{\rm K}$). Evidently, the proposed lower-limit $\log g$ will help to better characterize the two planet candidates detected in this system. The moderate observed level of activity for this relatively bright target ($m_{\rm Kep}\!<\!12.4$) does not alone explain the absence of detected oscillations\footnote{We should, however, note that the stellar inclination along the line of sight affects the apparent (i.e., observed) value of the magnetic activity proxy $\zeta_{\rm act}$, if we assume that the stellar variability in solar-type stars is dominated by contributions from active latitutes like for the Sun. Consequently, an intrinsically active star observed at a low angle of inclination may present a moderate-to-low activity proxy.}. In fact, there is a well-known high-temperature fall-off in the proportion of confirmed solar-like oscillators starting at $\sim\!6700\:{\rm K}$ \citep{VernerComparison}, in agreement with the location of the red edge of the classical instability strip. We attribute the absence of detected oscillations mainly to the latter effect.

\item KOI-100 (already highlighted in Sect.~\ref{sec:resnonosc}) has spectroscopic follow-up yielding $\log g\!=\!3.69\pm0.11\:{\rm dex}$. The non-detection of oscillations yields $\log g_{\rm md}\!=\!4.12^{+0.11}_{-0.10}\:{\rm dex}$, showing that the spectroscopic classification is erroneous and that the star is less evolved than previously assumed, likely an F-type star ($T_{\rm eff}\!=\!6743\pm140\:{\rm K}$) residing close to the main-sequence. The same reasons invoked above to explain the absence of detected oscillations in KOI-1525 apply here, to which we should add the target's faintness ($m_{\rm Kep}\!>\!12.4$). The computed $\log g_{\rm md}$ then suggests that the currently assumed size of the transiting object based on the spectroscopic solution ($\sim\!22\,R_\earth$) is too large, and that the companion is likely a genuine planet rather than a low-mass star or brown dwarf. Again, the proposed lower-limit $\log g$ clearly contributes to better characterizing the planet candidate.

\end{enumerate}

This work is an example of the enduring synergy between asteroseismology and exoplanetary science. Throughout the course of the {\it Kepler} mission, asteroseismology has played an important role in the characterization of planet-candidate host stars. Here, we give continuity to this effort by providing limits on stellar properties of planet-candidate host stars from the non-detection of solar-like oscillations.

\acknowledgments
{\it Kepler} was competitively selected as the tenth Discovery mission. Funding for this mission is provided by NASA's Science Mission Directorate. The authors wish to thank the entire {\it Kepler} team, without whom these results would not be possible. TLC, WJC, RH, AM, GRD and YPE acknowledge the support of the UK Science and Technology Facilities Council (STFC). The research leading to these results has received funding from the European CommunityÕs Seventh Framework Programme (FP7/2007--2013) under grant agreement no.~312844 (SPACEINN). Funding for the Stellar Astrophysics Centre is provided by The Danish National Research Foundation (Grant DNRF106). The research is supported by the ASTERISK project (ASTERoseismic Investigations with SONG and {\it Kepler}) funded by the European Research Council (Grant agreement no.: 267864). DH is supported by an appointment to the NASA Postdoctoral Program at Ames Research Center, administered by Oak Ridge Associated Universities through a contract with NASA. SH acknowledges financial support from the Netherlands Organization for Scientific Research (NWO). SH acknowledges support from the European Research Council under the European Community's Seventh Framework Programme (FP7/2007--2013)/ERC grant agreement no.~338251 (StellarAges). EC acknowledges financial support from the European Research Council under the European Community's Seventh Framework Programme (FP7/2007--2013)/ERC grant agreement no.~227224 (PROSPERITY), from the Fund for Scientific Research of Flanders (G.0728.11), from the Belgian federal science policy office (C90291 Gaia-DPAC), and from Project ASK under grant agreement no.~PIRSES-GA-2010-269194. SB acknowledges support from NSF grant AST-1105930 and NASA grant NNX13AE70G. TSM acknowledges NASA grant NNX13AE91G.

\appendix

\section{Computation of mode amplitude threshold $A_{\rm max,md}$}\label{sec:threshold}
\citet{ChaplinDetect} devised a statistical test for predicting the detectability of solar-like oscillations in any given {\it Kepler} target. The same test is used in this work, although employed in the reverse order. In other words, given the probability of detecting the oscillations, we may translate it into a global measure of the signal-to-noise ratio in the oscillation spectrum, ${\rm S/N_{tot}}$, required to make a marginal detection. Finally, knowledge of the noise background will make it possible to compute the mode amplitude threshold, $A_{\rm max,md}$, required for detection. The noise background is frequency-dependent, and so too will be $A_{\rm max,md}$, meaning that the statistical test must be applied at different frequencies. These frequencies can be regarded as proxies of $\nu_{\rm max}$. The steps involved in the computation of $A_{\rm max,md}$ are summarized next:
\begin{enumerate}

\item Computation of ${\rm S/N_{thresh}}$:
\begin{enumerate}
\item A total of $N$ independent frequency bins enter the estimation of ${\rm S/N_{tot}}$:
\begin{equation}
N = \nu_{\rm max,proxy}\,T \, ,
\label{eq:indbins}
\end{equation}
where $T$ is the length of the observations. We have assumed that the mode power is contained within a range $\pm\nu_{\rm max,proxy}/2$ around $\nu_{\rm max,proxy}$.
\item We begin by testing the $H_0$ or null hypothesis (i.e., that we observe pure noise). When binning over $N$ bins, the statistics of the power spectrum of a pure noise signal is taken to be $\chi^2$ with $2N$ degrees of freedom \citep[][]{AppourBin}. By specifying a $p$-value, we proceed with the numerical computation\footnote{A formula for the percent point function of the $\chi^2$ distribution does not exist in a simple closed form and hence it is computed numerically.} of the detection threshold ${\rm S/N_{thresh}}$:
\begin{equation}
p = \int_x^\infty \frac{\exp(-x')}{{\rm \Gamma}(N)} \, x'^{(N-1)} \, {\rm d}x' \, ,
\label{eq:snr_thresh}    
\end{equation} 
where $x\!=\!1+{\rm S/N_{thresh}}$ and ${\rm \Gamma}$ is the gamma function. According to Eqs.~(\ref{eq:indbins}) and (\ref{eq:snr_thresh}), an increase in $T$ results in a reduction in ${\rm S/N_{thresh}}$. Hence, for a given underlying ${\rm S/N_{tot}}$, the oscillation power will be more noticeable against the background as $T$ increases.
\end{enumerate}

\item Computation of ${\rm S/N_{tot}}$: The probability, $p_{{\rm detect}}$, that ${\rm S/N_{tot}}$ exceeds ${\rm S/N_{thresh}}$ is also given by Eq.~(\ref{eq:snr_thresh}), although by instead setting $x\!=\!(1+{\rm S/N_{thresh}})/(1+{\rm S/N_{tot}})$. This step can be thought of as testing the $H_1$ or alternative hypothesis (i.e., that we observe a signal embedded in noise). Having adopted a detection probability, $p_{{\rm detect}}$, we again have to rely on a numerical computation in order to obtain ${\rm S/N_{tot}}$. We adopted $p\!=\!0.01$ and $p_{\rm detect}\!=\!0.90$ throughout this work.

\item Computation of $A_{\rm max,md}$:
\begin{enumerate}
\item ${\rm S/N_{tot}}$ is given by the ratio of the total mean mode power, $P_{{\rm tot}}$, to the total background power across the frequency range occupied by the oscillations, $B_{{\rm tot}}$. The latter is approximately given by
\begin{equation}
B_{{\rm tot}} \approx b_{{\rm max}}\,\nu_{{\rm max,proxy}} \, ,
\label{eq:B_tot}
\end{equation}
where $b_{{\rm max}}$ is the background power-spectral density from granulation and instrumental/shot noise at $\nu_{{\rm max,proxy}}$. For a given data set we determined the background power-spectral density in one of two ways: \begin{inparaenum}[(i)] \item by fitting a Harvey-like profile plus a constant offset to the power spectrum when oscillations are present \citep[e.g.,][and references therein]{Faculae}; or \item by applying a median filter when no oscillations are present. \end{inparaenum}
\item The total mean mode power, $P_{{\rm tot}}$, may be approximately expressed as
\begin{equation}
P_{\rm tot} \approx 1.55\,A_{\rm max}^2\,\eta^2\,\frac{\nu_{\rm max,proxy}}{\Delta \nu} \, ,
\label{eq:P_tot}
\end{equation}
where $\eta$ takes into account the apodization of the oscillation signal due to the sampling. From Eqs.~(\ref{eq:B_tot}) and (\ref{eq:P_tot}), we obtain
\begin{equation}
A_{\rm max,md} = \left(\frac{1}{1.55\,\eta^2}\:\Delta \nu\:b_{\rm max}\:{\rm S/N_{tot}}\right)^{1/2} \, .  
\label{eq:amaxmd}    
\end{equation}
A value of $\Delta \nu$ consistent with $\nu_{\rm max,proxy}$ can be obtained from a scaling relation between $\Delta \nu$ and $\nu_{\rm max}$ \citep[e.g.,][]{StelloScale}. The computation of $A_{\rm max,md}$ incorporates the effect of the observing conditions, namely, the length of the observations via the quantity $T$ (Eq.~\ref{eq:indbins}) and the stellar magnitude via the quantity $b_{\rm max}$ (Eqs.~\ref{eq:B_tot} and \ref{eq:amaxmd}; since the instrumental/shot noise level is magnitude-dependent).
\end{enumerate}

\end{enumerate}

\section{Computation of magnetic activity proxy}\label{sec:actproxy}
The magnetic activity proxy, $\zeta_{\rm act}$, is simply an estimate of the intrinsic stellar noise and is intended to measure the level of activity of a star. This variability metric ultimately comprises contributions from rotational spot-modulation, chromospheric activity and stellar magnetic cycles. To compute it, we made use of {\it Kepler} long-cadence data \citep[$\Delta t\!\sim\!30\:{\rm min}$;][]{LCdata}, whose sampling cadence is adequate for the purpose of this calculation, since it is far exceeded by the typical timescales of the expected contributing factors. Specifically, the PDC (Presearch Data Conditioning) version of the data was used, since it has been corrected for systematic errors by the Science Operations Center (SOC) pipeline \citep[][]{PDC}.

The magnetic activity proxy was computed for all targets classified as solar-type stars in the framework of the KASC. This totaled 2750 targets. It should be stressed, however, that there is no guarantee that this sample is exclusively composed of bona fide solar-type stars. Furthermore, the proxy has been computed for an additional 885 KOIs, corresponding to the complete set of KOIs for which short-cadence data are available. Those KOIs for which the ephemerides of the candidate planet(s) are known \citep[see][]{BatalhaCandidates3Ephemerides} had their transit signals removed prior to the proxy estimation. This was done by removing segments of the time series equal to $1.5$ times the transit duration and centered on the time of mid transit. The long-cadence observations of the selected targets were taken from Q0 through Q14. No restrictions have been imposed in terms of magnitude or number of available quarters.

The proxy estimation was performed as follows:
\begin{enumerate}

\item For each target, we applied a binning of 11 data points ($\sim\!5.5\:{\rm hr}$) to each quarter of data\footnote{A binning of 11 data points translates into an effective cutoff of $0.43\:{\rm d}$ (3-dB bandwidth), meaning that one should get a sensible measure of the magnetic activity level even for the fastest rotators in the sample (with rotational periods $P_{\rm rot}\!\gtrsim\!0.5\:{\rm d}$).}. The proxy was then given, for each quarter ${\rm Q}_i$, by a constant scale factor $k$ times the median absolute deviation\footnote{For a univariate data set, the MAD is defined as the median of the absolute deviations from the data's median. The multiplicative constant is taken to be $k\!=\!1.4826$, thus converting the MAD into a consistent estimator of the standard deviation, under the assumption of normally distributed data.} (MAD) of the smoothed time series, i.e., $\zeta_{{\rm act,}{\rm Q}_i}\!=\!k \cdot {\rm MAD}_{{\rm Q}_i}$. The uncertainty in the quarterly proxy, $\zeta_{{\rm error,}{\rm Q}_i}$, was given by $\zeta_{{\rm act,}{\rm Q}_i}/\sqrt{2(N_i-1)}$, where $N_i$ is the number of data points in ${\rm Q}_i$. The reason for using the MAD relies on the fact that it is a robust measure of statistical dispersion, being less prone to outliers than the standard deviation about the mean. The latter has been used by \citet{starspotproxy} to define what they termed the starspot proxy.

\item A magnitude-dependent additive correction was applied to the quarterly proxy estimates to take into account the contribution due to instrumental/shot noise. To implement this correction, we used the minimal term model for the noise proposed by \citet{LCdata}, which gives the RMS noise, $\hat{\sigma}_{\rm lower}$, per integration. Since the time series were binned over $M$ points ($M\!=\!11$ in the present case), the additive correction to be removed from $\zeta_{{\rm act,}{\rm Q}_i}$ was just $\hat{\sigma}_{\rm lower}/\sqrt{M}$.

\item The magnetic activity proxy, $\zeta_{\rm act}$, was taken as the median of the quarterly proxy estimates, with an associated uncertainty ($\zeta_{\rm error}$) given by the MAD of those same quarterly estimates times the constant scale factor $k$.

\end{enumerate}

\begin{figure}
\epsscale{1.0}
\plotone{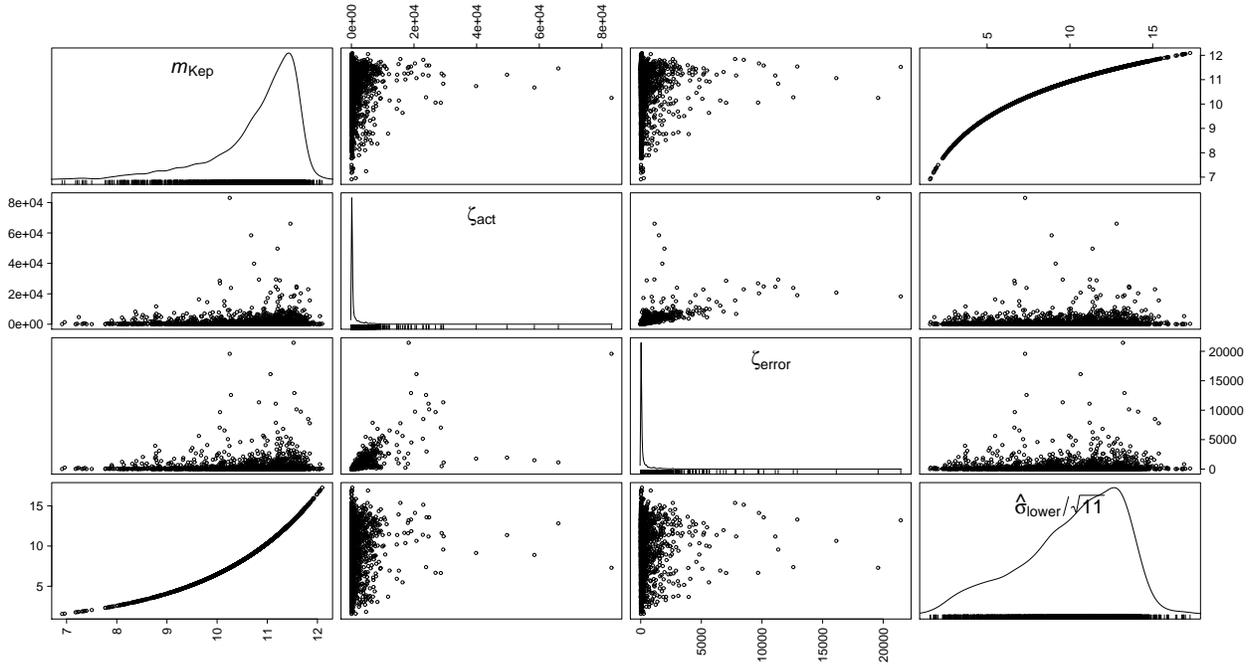}
\caption{\small Scatter plot matrix showing the results of the proxy estimation for the KASC targets. Parameters entering the plots are the {\it Kepler}-band magnitude ($m_{\rm Kep}$), the magnetic activity proxy ($\zeta_{\rm act}$), the uncertainty in the proxy ($\zeta_{\rm error}$) and the applied correction ($\hat{\sigma}_{\rm lower}/\sqrt{11}$). The curves in the diagonal panels represent the density distributions of the indicated parameters (ticks on the horizontal axis mark the position of each data point). Off-diagonal panels display all the possible pairwise correlations.\label{fig:scamatrix1}}
\end{figure}

\begin{figure}
\epsscale{1.0}
\plotone{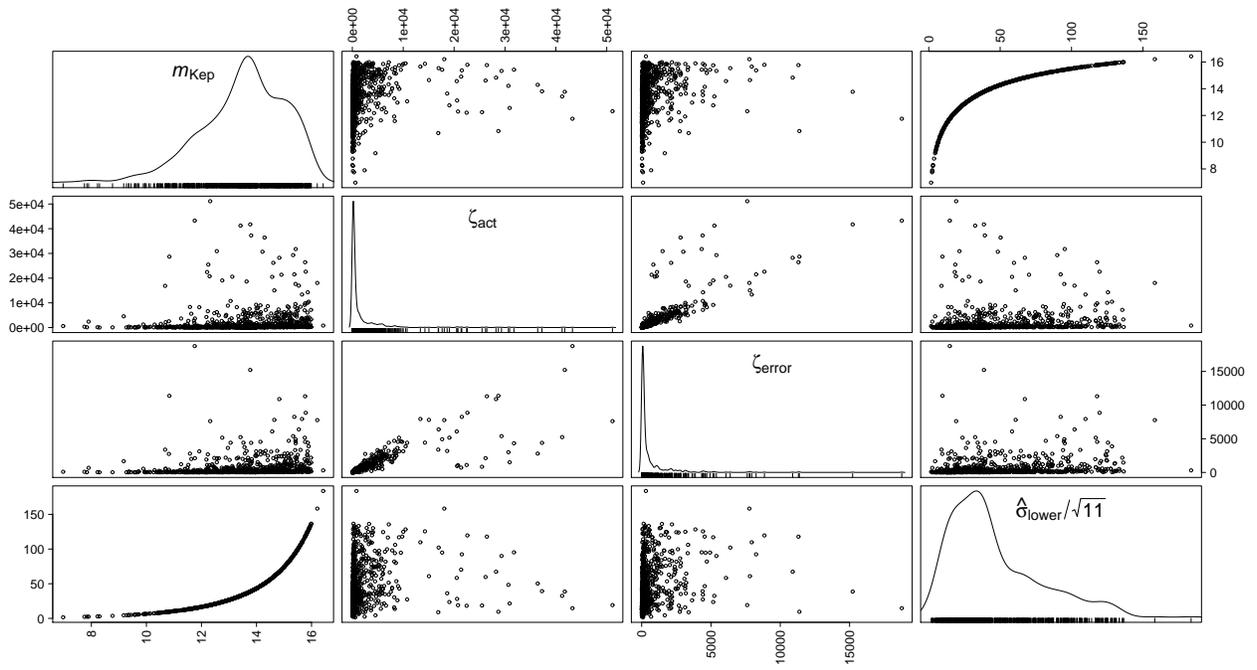}
\caption{\small Same as Fig.~\ref{fig:scamatrix1}, but for the KOIs.\label{fig:scamatrix2}}
\end{figure}

Figures \ref{fig:scamatrix1} and \ref{fig:scamatrix2} display the results of the proxy estimation in the form of scatter plot matrices for the KASC targets and KOIs, respectively. As desired, no correlation is seen between the magnitude and the value of the proxy. The same holds true between the applied correction and the proxy.


\bibliographystyle{apj}
\bibliography{biblio}

\clearpage


\begin{deluxetable}{rrccccr}
\tabletypesize{\scriptsize}
\tablewidth{0pt}
\tablenum{4}
\tablecaption{Proposed lower-limit $\log g$ estimates for solar-type KOIs with no detected oscillations.\label{tb:finaltable}}
\tablehead{\colhead{KOI\tablenotemark{a}}&\colhead{KIC}&\colhead{$m_{\rm Kep}$}&\colhead{$T_{\rm eff}$\tablenotemark{b}}&\colhead{$\zeta_{\rm act}$}&\colhead{$\log g_{\rm md}$\tablenotemark{c}}&\colhead{$\Delta \log g$} \\
\colhead{}&\colhead{}&\colhead{}&\colhead{(K)}&\colhead{(ppm)}&\colhead{(dex)}&\colhead{(dex)}}
\startdata
\nodata & 10909274 & 12.153 & 6594$\pm$93 & 456$\pm$376 & $4.11^{+0.22\:(0.21)}_{-0.18\:(0.17)}$ & \nodata \\
\nodata & 6131659 & 12.534 & 5087$\pm$63 & 797$\pm$590 & $3.78^{+0.33\:(0.33)}_{-0.23\:(0.23)}$ & \nodata \\
\nodata & 7677005 & 12.178 & 6901$\pm$94 & 1011$\pm$507 & $3.88^{+0.28\:(0.28)}_{-0.25\:(0.24)}$ & \nodata \\
\nodata & 4144236 & 11.856 & 6618$\pm$101 & 709$\pm$275 & $4.07^{+0.15\:(0.15)}_{-0.14\:(0.13)}$ & \nodata \\
\nodata & 6593363 & 12.893 & 6170$\pm$79 & 539$\pm$293 & $3.70^{+0.17\:(0.17)}_{-0.14\:(0.13)}$ & \nodata \\
\nodata & 5653126 & 13.173 & 6002$\pm$124 & 641$\pm$400 & $3.68^{+0.22\:(0.22)}_{-0.16\:(0.15)}$ & \nodata \\
\nodata & 6522750 & 11.230 & 5801$\pm$59 & 619$\pm$341 & $4.05^{+0.18\:(0.18)}_{-0.18\:(0.17)}$ & \nodata \\
\nodata & 2693092 & 12.003 & 5996$\pm$94 & 392$\pm$263 & $4.12^{+0.15\:(0.14)}_{-0.13\:(0.13)}$ & \nodata \\
4 & 3861595 & 11.432 & 6220$\pm$101 & 583$\pm$160 & $4.06^{+0.08\:(0.07)}_{-0.09\:(0.08)}$ & $-$0.25$\pm$0.09 \\
6 & 3248033 & 12.161 & 6558$\pm$80 & 195$\pm$57 & $3.95^{+0.05\:(0.03)}_{-0.05\:(0.03)}$ & \nodata \\
9 & 11553706 & 13.123 & 6288$\pm$72 & 280$\pm$191 & $3.71^{+0.11\:(0.10)}_{-0.09\:(0.08)}$ & \nodata \\
10 & 6922244 & 13.563 & 6392$\pm$82 & 175$\pm$51 & $3.83^{+0.05\:(0.03)}_{-0.05\:(0.04)}$ & 0.28$\pm$0.05 \\
11 & 11913073 & 13.496 & 5478$\pm$80 & 371$\pm$342 & $3.58^{+0.18\:(0.18)}_{-0.10\:(0.09)}$ & \nodata \\
12 & 5812701 & 11.353 & 6635$\pm$71 & 416$\pm$106 & $4.45^{+0.06\:(0.04)}_{-0.07\:(0.05)}$ & $-$0.19$\pm$0.06 \\
20 & 11804465 & 13.438 & 6279$\pm$89 & 238$\pm$70 & $3.87^{+0.05\:(0.03)}_{-0.06\:(0.04)}$ & 0.36$\pm$0.06 \\
21 & 10125352 & 13.396 & 6414$\pm$89 & 392$\pm$95 & $3.57^{+0.07\:(0.05)}_{-0.07\:(0.05)}$ & \nodata \\
22 & 9631995 & 13.435 & 6078$\pm$76 & 209$\pm$113 & $3.97^{+0.06\:(0.05)}_{-0.07\:(0.05)}$ & 0.44$\pm$0.07 \\
23 & 9071386 & 12.291 & 6540$\pm$81 & 552$\pm$72 & $3.81^{+0.06\:(0.04)}_{-0.06\:(0.05)}$ & \nodata \\
70 & 6850504 & 12.498 & 5540$\pm$63 & 455$\pm$137 & $4.17^{+0.08\:(0.07)}_{-0.08\:(0.07)}$ & 0.28$\pm$0.08 \\
78 & 9764820 & 10.870 & 5390$\pm$57 & 244$\pm$104 & $4.32^{+0.07\:(0.06)}_{-0.06\:(0.04)}$ & \nodata \\
84 & 2571238 & 11.898 & 5623$\pm$57 & 120$\pm$70 & $4.40^{+0.05\:(0.04)}_{-0.05\:(0.03)}$ & 0.13$\pm$0.05 \\
88 & 7700871 & 11.871 & 5801$\pm$63 & 77$\pm$44 & $4.33^{+0.05\:(0.03)}_{-0.05\:(0.02)}$ & \nodata \\
91 & 7747867 & 11.684 & 5834$\pm$63 & 192$\pm$101 & $4.14^{+0.08\:(0.07)}_{-0.07\:(0.05)}$ & \nodata \\
92 & 7941200 & 11.667 & 6069$\pm$69 & 334$\pm$124 & $4.39^{+0.07\:(0.05)}_{-0.07\:(0.06)}$ & 0.04$\pm$0.07 \\
93 & 6784857 & 11.477 & 6279$\pm$62 & 124$\pm$11 & $4.27^{+0.04\:(0.01)}_{-0.04\:(0.01)}$ & \nodata \\
99 & 8505215 & 12.960 & 5162$\pm$63 & 230$\pm$101 & $4.16^{+0.06\:(0.05)}_{-0.07\:(0.05)}$ & 0.46$\pm$0.06 \\
100 & 4055765 & 12.598 & 6743$\pm$140 & 399$\pm$183 & $4.12^{+0.11\:(0.10)}_{-0.10\:(0.09)}$ & $-$0.43$\pm$0.10 \\
102 & 8456679 & 12.566 & 6242$\pm$138 & 186$\pm$156 & $4.25^{+0.08\:(0.06)}_{-0.09\:(0.08)}$ & 0.17$\pm$0.08 \\
103 & 2444412 & 12.593 & 5766$\pm$63 & 835$\pm$280 & $3.87^{+0.16\:(0.16)}_{-0.15\:(0.15)}$ & 0.57$\pm$0.16 \\
105 & 8711794 & 12.870 & 5809$\pm$101 & 75$\pm$38 & $4.18^{+0.04\:(0.02)}_{-0.04\:(0.02)}$ & 0.22$\pm$0.04 \\
106 & 10489525 & 12.775 & 6730$\pm$133 & 176$\pm$24 & $3.81^{+0.04\:(0.01)}_{-0.04\:(0.01)}$ & \nodata \\
109 & 4752451 & 12.385 & 6201$\pm$79 & 115$\pm$35 & $4.15^{+0.04\:(0.02)}_{-0.04\:(0.02)}$ & \nodata \\
110 & 9450647 & 12.663 & 6538$\pm$79 & 52$\pm$22 & $4.19^{+0.04\:(0.01)}_{-0.04\:(0.01)}$ & 0.17$\pm$0.04 \\
111 & 6678383 & 12.596 & 6170$\pm$65 & 24$\pm$13 & $4.31^{+0.04\:(0.01)}_{-0.04\:(0.01)}$ & 0.10$\pm$0.04 \\
112 & 10984090 & 12.772 & 6125$\pm$73 & 83$\pm$57 & $4.10^{+0.05\:(0.03)}_{-0.05\:(0.03)}$ & 0.36$\pm$0.05 \\
114 & 6721123 & 12.660 & 6365$\pm$86 & 140$\pm$28 & $3.86^{+0.04\:(0.01)}_{-0.04\:(0.01)}$ & \nodata \\
115 & 9579641 & 12.791 & 6397$\pm$91 & 175$\pm$78 & $4.10^{+0.06\:(0.04)}_{-0.06\:(0.04)}$ & 0.14$\pm$0.06 \\
116 & 8395660 & 12.882 & 6280$\pm$109 & 146$\pm$52 & $4.10^{+0.05\:(0.03)}_{-0.05\:(0.03)}$ & 0.31$\pm$0.05 \\
120 & 11869052 & 12.003 & 5632$\pm$273 & 188$\pm$34 & $4.05^{+0.04\:(0.02)}_{-0.05\:(0.03)}$ & \nodata \\
121 & 3247396 & 12.759 & 6390$\pm$85 & 107$\pm$76 & $3.99^{+0.05\:(0.03)}_{-0.06\:(0.04)}$ & \nodata \\
124 & 11086270 & 12.935 & 6314$\pm$83 & 37$\pm$30 & $4.20^{+0.04\:(0.02)}_{-0.04\:(0.01)}$ & $-$0.13$\pm$0.04 \\
128 & 11359879 & 13.758 & 5841$\pm$132 & 369$\pm$208 & $3.67^{+0.13\:(0.12)}_{-0.13\:(0.12)}$ & 0.75$\pm$0.13 \\
129 & 11974540 & 13.224 & 6741$\pm$108 & 328$\pm$73 & $3.76^{+0.06\:(0.04)}_{-0.05\:(0.04)}$ & \nodata \\
130 & 5297298 & 13.325 & 6237$\pm$105 & 132$\pm$70 & $3.90^{+0.05\:(0.04)}_{-0.05\:(0.04)}$ & \nodata \\
132 & 8892910 & 13.794 & 6176$\pm$103 & 213$\pm$34 & $3.60^{+0.05\:(0.03)}_{-0.05\:(0.03)}$ & \nodata \\
134 & 9032900 & 13.675 & 6357$\pm$106 & 226$\pm$49 & $3.79^{+0.05\:(0.03)}_{-0.05\:(0.03)}$ & \nodata \\
137 & 8644288 & 13.549 & 5394$\pm$91 & 163$\pm$74 & $3.97^{+0.06\:(0.04)}_{-0.05\:(0.04)}$ & 0.46$\pm$0.06 \\
139 & 8559644 & 13.492 & 6145$\pm$94 & 302$\pm$152 & $3.93^{+0.08\:(0.07)}_{-0.09\:(0.08)}$ & 0.45$\pm$0.08 \\
141 & 12105051 & 13.687 & 5402$\pm$89 & 481$\pm$144 & $3.65^{+0.08\:(0.07)}_{-0.15\:(0.14)}$ & 0.85$\pm$0.12 \\
142 & 5446285 & 13.113 & 5559$\pm$79 & 449$\pm$414 & $4.00^{+0.20\:(0.20)}_{-0.21\:(0.20)}$ & 0.47$\pm$0.21 \\
143 & 4649305 & 13.682 & 6984$\pm$130 & 174$\pm$47 & $3.85^{+0.05\:(0.03)}_{-0.05\:(0.03)}$ & \nodata \\
146 & 9048161 & 13.030 & 6302$\pm$114 & 151$\pm$91 & $3.92^{+0.06\:(0.05)}_{-0.07\:(0.05)}$ & \nodata \\
148 & 5735762 & 13.040 & 5189$\pm$71 & 530$\pm$125 & $3.87^{+0.08\:(0.07)}_{-0.08\:(0.07)}$ & 0.62$\pm$0.08 \\
149 & 3835670 & 13.397 & 5718$\pm$88 & 102$\pm$39 & $3.92^{+0.04\:(0.02)}_{-0.05\:(0.02)}$ & 0.05$\pm$0.05 \\
150 & 7626506 & 13.771 & 5822$\pm$91 & 160$\pm$75 & $3.92^{+0.05\:(0.04)}_{-0.06\:(0.04)}$ & 0.54$\pm$0.05 \\
152 & 8394721 & 13.914 & 6405$\pm$103 & 169$\pm$107 & $3.80^{+0.07\:(0.05)}_{-0.07\:(0.06)}$ & 0.68$\pm$0.07 \\
154 & 9970525 & 13.174 & 6510$\pm$97 & 125$\pm$43 & $3.76^{+0.05\:(0.03)}_{-0.05\:(0.03)}$ & \nodata \\
155 & 8030148 & 13.494 & 5954$\pm$91 & 75$\pm$54 & $3.92^{+0.05\:(0.03)}_{-0.05\:(0.03)}$ & 0.27$\pm$0.05 \\
157 & 6541920 & 13.709 & 5919$\pm$95 & 112$\pm$69 & $3.98^{+0.05\:(0.03)}_{-0.05\:(0.04)}$ & 0.40$\pm$0.05 \\
159 & 8972058 & 13.431 & 6069$\pm$91 & 176$\pm$34 & $3.87^{+0.05\:(0.02)}_{-0.05\:(0.02)}$ & 0.44$\pm$0.05 \\
160 & 6631721 & 13.805 & 6405$\pm$115 & 119$\pm$48 & $3.77^{+0.05\:(0.03)}_{-0.05\:(0.03)}$ & \nodata \\
162 & 8107380 & 13.837 & 5817$\pm$95 & 120$\pm$51 & $3.81^{+0.05\:(0.02)}_{-0.05\:(0.03)}$ & 0.66$\pm$0.05 \\
163 & 6851425 & 13.536 & 5264$\pm$70 & 157$\pm$64 & $3.87^{+0.05\:(0.04)}_{-0.05\:(0.03)}$ & 0.62$\pm$0.05 \\
165 & 9527915 & 13.938 & 5201$\pm$63 & 293$\pm$57 & $3.67^{+0.05\:(0.03)}_{-0.05\:(0.03)}$ & 0.86$\pm$0.05 \\
166 & 2441495 & 13.575 & 5386$\pm$83 & 306$\pm$131 & $3.71^{+0.08\:(0.07)}_{-0.08\:(0.07)}$ & 0.70$\pm$0.08 \\
167 & 11666881 & 13.273 & 6485$\pm$78 & 93$\pm$44 & $3.92^{+0.05\:(0.02)}_{-0.05\:(0.03)}$ & 0.49$\pm$0.05 \\
169 & 6185711 & 13.579 & 5815$\pm$68 & 240$\pm$140 & $3.61^{+0.08\:(0.07)}_{-0.09\:(0.08)}$ & \nodata \\
171 & 7831264 & 13.717 & 6495$\pm$105 & 105$\pm$44 & $3.86^{+0.05\:(0.02)}_{-0.05\:(0.02)}$ & 0.54$\pm$0.05 \\
172 & 8692861 & 13.749 & 5886$\pm$69 & 98$\pm$93 & $3.86^{+0.07\:(0.06)}_{-0.06\:(0.05)}$ & 0.80$\pm$0.07 \\
173 & 11402995 & 13.844 & 6030$\pm$78 & 72$\pm$43 & $3.84^{+0.05\:(0.02)}_{-0.05\:(0.02)}$ & 0.45$\pm$0.05 \\
175 & 8323753 & 13.488 & 6078$\pm$95 & 147$\pm$32 & $3.85^{+0.04\:(0.02)}_{-0.04\:(0.02)}$ & \nodata \\
176 & 6442377 & 13.432 & 6568$\pm$93 & 56$\pm$39 & $3.98^{+0.04\:(0.02)}_{-0.05\:(0.02)}$ & 0.45$\pm$0.04 \\
177 & 6803202 & 13.182 & 5870$\pm$79 & 93$\pm$79 & $3.98^{+0.05\:(0.04)}_{-0.06\:(0.05)}$ & 0.40$\pm$0.06 \\
179 & 9663113 & 13.955 & 6081$\pm$90 & 70$\pm$16 & $3.75^{+0.04\:(0.01)}_{-0.04\:(0.01)}$ & 0.69$\pm$0.04 \\
192 & 7950644 & 14.221 & 6195$\pm$89 & 117$\pm$89 & $3.67^{+0.06\:(0.05)}_{-0.07\:(0.05)}$ & 0.79$\pm$0.06 \\
198 & 10666242 & 14.288 & 5736$\pm$85 & 108$\pm$33 & $3.63^{+0.05\:(0.03)}_{-0.06\:(0.04)}$ & \nodata \\
200 & 6046540 & 14.412 & 5945$\pm$119 & 212$\pm$94 & $3.55^{+0.07\:(0.06)}_{-0.07\:(0.06)}$ & 0.96$\pm$0.07 \\
201 & 6849046 & 14.014 & 5629$\pm$114 & 269$\pm$158 & $3.68^{+0.10\:(0.09)}_{-0.08\:(0.07)}$ & 0.64$\pm$0.09 \\
209 & 10723750 & 14.274 & 6438$\pm$103 & 217$\pm$46 & $3.75^{+0.05\:(0.03)}_{-0.05\:(0.03)}$ & 0.68$\pm$0.05 \\
232 & 4833421 & 14.247 & 6102$\pm$84 & 82$\pm$74 & $3.81^{+0.06\:(0.04)}_{-0.06\:(0.04)}$ & 0.72$\pm$0.06 \\
238 & 7219825 & 14.061 & 6274$\pm$103 & 44$\pm$18 & $3.65^{+0.04\:(0.01)}_{-0.04\:(0.01)}$ & 0.80$\pm$0.04 \\
241 & 11288051 & 14.139 & 5288$\pm$60 & 119$\pm$56 & $3.53^{+0.05\:(0.03)}_{-0.05\:(0.03)}$ & 1.19$\pm$0.05 \\
258 & 11231334 & 9.887 & 6528$\pm$91 & 1242$\pm$550 & $4.35^{+0.25\:(0.25)}_{-0.37\:(0.37)}$ & $-$0.18$\pm$0.31 \\
259 & 5790807 & 9.954 & 6581$\pm$57 & 362$\pm$100 & $4.50^{+0.05\:(0.02)}_{-0.06\:(0.05)}$ & \nodata \\
261 & 5383248 & 10.297 & 5779$\pm$66 & 1209$\pm$337 & $4.16^{+0.19\:(0.19)}_{-0.18\:(0.17)}$ & 0.26$\pm$0.18 \\
265 & 12024120 & 11.994 & 6277$\pm$62 & 109$\pm$46 & $4.21^{+0.05\:(0.02)}_{-0.05\:(0.02)}$ & 0.15$\pm$0.05 \\
283 & 5695396 & 11.525 & 5875$\pm$84 & 723$\pm$268 & $4.24^{+0.14\:(0.14)}_{-0.13\:(0.13)}$ & 0.18$\pm$0.14 \\
284 & 6021275 & 11.818 & 6176$\pm$73 & 57$\pm$32 & $4.27^{+0.04\:(0.01)}_{-0.04\:(0.01)}$ & 0.13$\pm$0.04 \\
291 & 10933561 & 12.848 & 5685$\pm$63 & 93$\pm$36 & $3.87^{+0.05\:(0.02)}_{-0.05\:(0.03)}$ & 0.32$\pm$0.05 \\
292 & 11075737 & 12.872 & 6025$\pm$73 & 209$\pm$90 & $4.07^{+0.06\:(0.04)}_{-0.06\:(0.05)}$ & 0.36$\pm$0.06 \\
294 & 11259686 & 12.674 & 6125$\pm$66 & 244$\pm$70 & $3.83^{+0.05\:(0.03)}_{-0.05\:(0.04)}$ & 0.61$\pm$0.05 \\
296 & 11802615 & 12.935 & 6088$\pm$68 & 80$\pm$35 & $3.92^{+0.05\:(0.02)}_{-0.04\:(0.02)}$ & 0.49$\pm$0.04 \\
297 & 11905011 & 12.182 & 6282$\pm$66 & 157$\pm$28 & $4.23^{+0.04\:(0.02)}_{-0.04\:(0.01)}$ & 0.04$\pm$0.04 \\
301 & 3642289 & 12.730 & 6337$\pm$93 & 58$\pm$30 & $3.97^{+0.04\:(0.01)}_{-0.04\:(0.02)}$ & 0.37$\pm$0.04 \\
302 & 3662838 & 12.059 & 6953$\pm$103 & 240$\pm$48 & $4.20^{+0.05\:(0.03)}_{-0.05\:(0.03)}$ & $-$0.07$\pm$0.05 \\
303 & 5966322 & 12.193 & 5734$\pm$60 & 364$\pm$153 & $4.09^{+0.09\:(0.08)}_{-0.09\:(0.08)}$ & 0.28$\pm$0.09 \\
304 & 6029239 & 12.549 & 6150$\pm$81 & 767$\pm$177 & $3.96^{+0.09\:(0.08)}_{-0.10\:(0.09)}$ & 0.05$\pm$0.10 \\
307 & 6289257 & 12.797 & 6310$\pm$76 & 188$\pm$86 & $4.02^{+0.06\:(0.04)}_{-0.06\:(0.04)}$ & 0.40$\pm$0.06 \\
308 & 6291837 & 12.351 & 6355$\pm$83 & 264$\pm$58 & $4.25^{+0.05\:(0.02)}_{-0.05\:(0.02)}$ & 0.12$\pm$0.05 \\
313 & 7419318 & 12.990 & 5331$\pm$79 & 184$\pm$96 & $4.03^{+0.06\:(0.05)}_{-0.06\:(0.05)}$ & 0.50$\pm$0.06 \\
316 & 8008067 & 12.701 & 5705$\pm$91 & 244$\pm$78 & $4.07^{+0.06\:(0.04)}_{-0.05\:(0.04)}$ & 0.26$\pm$0.06 \\
317 & 8121310 & 12.885 & 6658$\pm$126 & 111$\pm$48 & $4.13^{+0.05\:(0.03)}_{-0.05\:(0.03)}$ & $-$0.09$\pm$0.05 \\
318 & 8156120 & 12.211 & 6578$\pm$78 & 1133$\pm$181 & $3.85^{+0.13\:(0.12)}_{-0.11\:(0.11)}$ & 0.44$\pm$0.12 \\
321 & 8753657 & 12.520 & 5611$\pm$57 & 71$\pm$37 & $4.27^{+0.04\:(0.02)}_{-0.04\:(0.02)}$ & 0.07$\pm$0.04 \\
327 & 9881662 & 12.996 & 6354$\pm$77 & 177$\pm$64 & $3.86^{+0.05\:(0.04)}_{-0.05\:(0.03)}$ & 0.53$\pm$0.05 \\
328 & 9895004 & 12.820 & 5821$\pm$69 & 746$\pm$186 & $3.73^{+0.11\:(0.10)}_{-0.12\:(0.11)}$ & \nodata \\
329 & 10031885 & 13.478 & 6036$\pm$75 & 112$\pm$81 & $3.65^{+0.05\:(0.03)}_{-0.06\:(0.04)}$ & \nodata \\
331 & 10285631 & 13.497 & 5555$\pm$63 & 482$\pm$379 & $3.52^{+0.24\:(0.23)}_{-0.06\:(0.04)}$ & 0.58$\pm$0.15 \\
332 & 10290666 & 13.046 & 5756$\pm$65 & 68$\pm$44 & $3.86^{+0.04\:(0.02)}_{-0.05\:(0.03)}$ & 0.47$\pm$0.05 \\
339 & 10587105 & 13.763 & 6277$\pm$73 & 202$\pm$90 & $3.58^{+0.07\:(0.05)}_{-0.07\:(0.05)}$ & 0.95$\pm$0.07 \\
343 & 10982872 & 13.203 & 5945$\pm$75 & 315$\pm$153 & $3.83^{+0.09\:(0.08)}_{-0.09\:(0.08)}$ & 0.49$\pm$0.09 \\
344 & 11015108 & 13.400 & 5984$\pm$81 & 301$\pm$212 & $3.65^{+0.12\:(0.11)}_{-0.12\:(0.12)}$ & 0.72$\pm$0.12 \\
351 & 11442793 & 13.804 & 6329$\pm$86 & 216$\pm$147 & $3.70^{+0.09\:(0.08)}_{-0.09\:(0.08)}$ & 0.76$\pm$0.09 \\
354 & 11568987 & 13.235 & 6282$\pm$115 & 683$\pm$186 & $3.73^{+0.11\:(0.10)}_{-0.15\:(0.14)}$ & 0.71$\pm$0.13 \\
365 & 11623629 & 11.195 & 5611$\pm$57 & 221$\pm$246 & $4.43^{+0.11\:(0.10)}_{-0.12\:(0.11)}$ & 0.06$\pm$0.12 \\
367 & 4815520 & 11.105 & 5864$\pm$60 & 287$\pm$78 & $4.32^{+0.05\:(0.04)}_{-0.05\:(0.04)}$ & 0.04$\pm$0.05 \\
369 & 7175184 & 11.992 & 6377$\pm$60 & 106$\pm$33 & $3.95^{+0.04\:(0.01)}_{-0.05\:(0.02)}$ & 0.54$\pm$0.04 \\
373 & 7364176 & 12.765 & 6121$\pm$75 & 110$\pm$29 & $3.69^{+0.04\:(0.01)}_{-0.04\:(0.02)}$ & 0.76$\pm$0.04 \\
383 & 3342463 & 13.109 & 6411$\pm$88 & 124$\pm$130 & $3.63^{+0.08\:(0.06)}_{-0.08\:(0.07)}$ & \nodata \\
403 & 4247092 & 14.169 & 5784$\pm$82 & 101$\pm$54 & $3.57^{+0.05\:(0.03)}_{-0.05\:(0.04)}$ & 0.89$\pm$0.05 \\
405 & 5003117 & 14.026 & 5577$\pm$78 & 194$\pm$159 & $3.57^{+0.09\:(0.08)}_{-0.09\:(0.08)}$ & \nodata \\
416 & 6508221 & 14.290 & 5249$\pm$72 & 90$\pm$54 & $3.56^{+0.06\:(0.05)}_{-0.06\:(0.05)}$ & 1.09$\pm$0.06 \\
506 & 5780715 & 14.731 & 6021$\pm$121 & 199$\pm$103 & $3.50^{+0.08\:(0.07)}_{-0.04\:(0.01)}$ & 0.99$\pm$0.06 \\
508 & 6266741 & 14.387 & 5612$\pm$108 & 202$\pm$101 & $3.52^{+0.07\:(0.06)}_{-0.06\:(0.04)}$ & 0.93$\pm$0.06 \\
518 & 8017703 & 14.287 & 5037$\pm$63 & 226$\pm$103 & $3.47^{+0.09\:(0.08)}_{-0.04\:(0.02)}$ & 1.18$\pm$0.07 \\
528 & 9941859 & 14.598 & 5674$\pm$84 & 79$\pm$47 & $3.49^{+0.06\:(0.04)}_{-0.04\:(0.01)}$ & 0.91$\pm$0.05 \\
567 & 7445445 & 14.338 & 5817$\pm$85 & 202$\pm$81 & $3.64^{+0.06\:(0.04)}_{-0.07\:(0.06)}$ & 0.89$\pm$0.07 \\
568 & 7595157 & 14.140 & 5390$\pm$81 & 203$\pm$50 & $3.72^{+0.05\:(0.03)}_{-0.05\:(0.03)}$ & 0.87$\pm$0.05 \\
584 & 9146018 & 14.129 & 5524$\pm$67 & 163$\pm$89 & $3.67^{+0.06\:(0.04)}_{-0.06\:(0.05)}$ & 0.95$\pm$0.06 \\
591 & 9886221 & 14.396 & 5693$\pm$75 & 59$\pm$37 & $3.56^{+0.05\:(0.03)}_{-0.05\:(0.02)}$ & \nodata \\
611 & 6309763 & 14.022 & 6357$\pm$102 & 245$\pm$66 & $3.63^{+0.06\:(0.04)}_{-0.06\:(0.05)}$ & 0.83$\pm$0.06 \\
612 & 6587002 & 14.157 & 5231$\pm$99 & 140$\pm$76 & $3.67^{+0.06\:(0.04)}_{-0.06\:(0.04)}$ & 0.36$\pm$0.06 \\
625 & 4449034 & 13.592 & 6464$\pm$124 & 278$\pm$52 & $3.86^{+0.05\:(0.03)}_{-0.05\:(0.03)}$ & 0.01$\pm$0.05 \\
633 & 4841374 & 13.871 & 6070$\pm$113 & 65$\pm$15 & $3.51^{+0.04\:(0.01)}_{-0.04\:(0.02)}$ & 0.51$\pm$0.04 \\
645 & 5374854 & 13.716 & 6306$\pm$115 & 174$\pm$95 & $3.69^{+0.06\:(0.05)}_{-0.06\:(0.05)}$ & 0.40$\pm$0.06 \\
649 & 5613330 & 13.310 & 6288$\pm$102 & 101$\pm$43 & $3.55^{+0.06\:(0.04)}_{-0.05\:(0.03)}$ & 0.75$\pm$0.05 \\
653 & 5893123 & 13.858 & 6615$\pm$136 & 61$\pm$30 & $3.68^{+0.04\:(0.02)}_{-0.05\:(0.02)}$ & \nodata \\
655 & 5966154 & 13.004 & 6463$\pm$86 & 52$\pm$34 & $3.87^{+0.04\:(0.02)}_{-0.04\:(0.02)}$ & 0.45$\pm$0.04 \\
659 & 6125481 & 13.413 & 6721$\pm$125 & 135$\pm$98 & $3.55^{+0.08\:(0.06)}_{-0.05\:(0.03)}$ & 0.70$\pm$0.06 \\
660 & 6267535 & 13.532 & 5480$\pm$74 & 76$\pm$56 & $3.77^{+0.05\:(0.03)}_{-0.05\:(0.03)}$ & 0.18$\pm$0.05 \\
662 & 6365156 & 13.336 & 6148$\pm$85 & 60$\pm$35 & $3.60^{+0.04\:(0.02)}_{-0.04\:(0.02)}$ & 0.80$\pm$0.04 \\
664 & 6442340 & 13.484 & 5985$\pm$89 & 148$\pm$133 & $3.52^{+0.09\:(0.08)}_{-0.05\:(0.03)}$ & 0.76$\pm$0.07 \\
665 & 6685609 & 13.182 & 6080$\pm$87 & 470$\pm$131 & $3.75^{+0.08\:(0.07)}_{-0.08\:(0.07)}$ & 0.63$\pm$0.08 \\
672 & 7115785 & 13.998 & 5760$\pm$103 & 621$\pm$215 & $3.50^{+0.19\:(0.18)}_{-0.04\:(0.02)}$ & 0.91$\pm$0.12 \\
680 & 7529266 & 13.643 & 6327$\pm$94 & 140$\pm$125 & $3.70^{+0.07\:(0.06)}_{-0.13\:(0.12)}$ & 0.65$\pm$0.10 \\
681 & 7598128 & 13.204 & 6549$\pm$96 & 335$\pm$80 & $3.52^{+0.07\:(0.06)}_{-0.04\:(0.01)}$ & \nodata \\
692 & 8557374 & 13.648 & 5806$\pm$75 & 55$\pm$26 & $3.77^{+0.04\:(0.02)}_{-0.04\:(0.02)}$ & 0.80$\pm$0.04 \\
693 & 8738735 & 13.949 & 6352$\pm$84 & 187$\pm$97 & $3.57^{+0.06\:(0.05)}_{-0.07\:(0.06)}$ & 0.94$\pm$0.07 \\
695 & 8805348 & 13.437 & 6226$\pm$80 & 66$\pm$78 & $3.60^{+0.06\:(0.05)}_{-0.06\:(0.05)}$ & 0.77$\pm$0.06 \\
696 & 8869680 & 13.357 & 5966$\pm$136 & 78$\pm$33 & $3.85^{+0.04\:(0.02)}_{-0.04\:(0.01)}$ & \nodata \\
700 & 8962094 & 13.580 & 5922$\pm$84 & 42$\pm$46 & $3.98^{+0.05\:(0.02)}_{-0.05\:(0.02)}$ & 0.49$\pm$0.05 \\
701 & 9002278 & 13.725 & 5036$\pm$66 & 323$\pm$74 & $3.76^{+0.06\:(0.04)}_{-0.06\:(0.04)}$ & 0.93$\pm$0.06 \\
707 & 9458613 & 13.988 & 6212$\pm$94 & 92$\pm$58 & $3.85^{+0.05\:(0.03)}_{-0.05\:(0.03)}$ & 0.18$\pm$0.05 \\
708 & 9530945 & 13.998 & 6277$\pm$88 & 147$\pm$49 & $3.59^{+0.06\:(0.05)}_{-0.06\:(0.04)}$ & 0.88$\pm$0.06 \\
711 & 9597345 & 13.967 & 5612$\pm$103 & 122$\pm$35 & $3.80^{+0.04\:(0.02)}_{-0.05\:(0.02)}$ & 0.60$\pm$0.04 \\
718 & 9884104 & 13.764 & 6029$\pm$88 & 499$\pm$143 & $3.56^{+0.13\:(0.13)}_{-0.08\:(0.06)}$ & 0.71$\pm$0.10 \\
968 & 3560301 & 10.963 & 6962$\pm$87 & 151$\pm$41 & $4.19^{+0.04\:(0.02)}_{-0.04\:(0.02)}$ & \nodata \\
978 & 11494130 & 10.988 & 6673$\pm$107 & 304$\pm$126 & $4.06^{+0.08\:(0.07)}_{-0.08\:(0.07)}$ & \nodata \\
987 & 7295235 & 12.550 & 5482$\pm$76 & 1286$\pm$384 & $3.57^{+0.21\:(0.21)}_{-0.10\:(0.09)}$ & 0.93$\pm$0.15 \\
991 & 10154388 & 13.581 & 5938$\pm$106 & 113$\pm$70 & $3.82^{+0.06\:(0.04)}_{-0.06\:(0.04)}$ & 0.21$\pm$0.06 \\
1001 & 1871056 & 13.038 & 6235$\pm$118 & 225$\pm$123 & $3.92^{+0.07\:(0.06)}_{-0.08\:(0.07)}$ & $-$0.12$\pm$0.08 \\
1020 & 2309719 & 12.899 & 6059$\pm$85 & 245$\pm$66 & $3.95^{+0.05\:(0.04)}_{-0.05\:(0.04)}$ & 0.17$\pm$0.05 \\
1057 & 6066416 & 11.558 & 6806$\pm$93 & 106$\pm$50 & $4.00^{+0.05\:(0.03)}_{-0.05\:(0.03)}$ & \nodata \\
1113 & 2854914 & 13.703 & 6314$\pm$115 & 219$\pm$217 & $3.55^{+0.11\:(0.11)}_{-0.06\:(0.04)}$ & 0.83$\pm$0.09 \\
1116 & 2849805 & 13.333 & 6029$\pm$94 & 151$\pm$113 & $3.57^{+0.07\:(0.06)}_{-0.07\:(0.05)}$ & 0.83$\pm$0.07 \\
1128 & 6362874 & 13.507 & 5485$\pm$63 & 144$\pm$81 & $3.58^{+0.06\:(0.04)}_{-0.06\:(0.04)}$ & 0.94$\pm$0.06 \\
1150 & 8278371 & 13.326 & 5915$\pm$73 & 257$\pm$99 & $3.52^{+0.06\:(0.05)}_{-0.05\:(0.03)}$ & 0.85$\pm$0.06 \\
1151 & 8280511 & 13.404 & 5759$\pm$70 & 116$\pm$74 & $3.68^{+0.05\:(0.03)}_{-0.05\:(0.04)}$ & 0.94$\pm$0.05 \\
1162 & 10528068 & 12.783 & 6138$\pm$82 & 32$\pm$18 & $3.81^{+0.04\:(0.01)}_{-0.04\:(0.01)}$ & 0.46$\pm$0.04 \\
1169 & 10319385 & 13.248 & 5956$\pm$77 & 80$\pm$39 & $3.56^{+0.05\:(0.02)}_{-0.04\:(0.02)}$ & 0.90$\pm$0.04 \\
1175 & 10350571 & 13.290 & 5650$\pm$65 & 309$\pm$102 & $3.53^{+0.07\:(0.06)}_{-0.06\:(0.05)}$ & 0.50$\pm$0.07 \\
1185 & 3443790 & 11.840 & 6276$\pm$63 & 610$\pm$91 & $3.70^{+0.08\:(0.07)}_{-0.06\:(0.05)}$ & \nodata \\
1215 & 3939150 & 13.420 & 6050$\pm$94 & 106$\pm$92 & $3.87^{+0.06\:(0.05)}_{-0.06\:(0.05)}$ & 0.21$\pm$0.06 \\
1218 & 3442055 & 13.331 & 5870$\pm$85 & 109$\pm$83 & $3.64^{+0.06\:(0.04)}_{-0.06\:(0.04)}$ & 0.73$\pm$0.06 \\
1220 & 4043190 & 12.988 & 5163$\pm$68 & 319$\pm$241 & $3.53^{+0.16\:(0.15)}_{-0.08\:(0.07)}$ & 0.48$\pm$0.12 \\
1236 & 6677841 & 13.659 & 6779$\pm$103 & 251$\pm$62 & $3.74^{+0.05\:(0.04)}_{-0.05\:(0.03)}$ & 0.71$\pm$0.05 \\
1242 & 6607447 & 13.750 & 6446$\pm$120 & 81$\pm$61 & $3.57^{+0.05\:(0.03)}_{-0.05\:(0.03)}$ & 0.86$\pm$0.05 \\
1275 & 8583696 & 13.672 & 5625$\pm$77 & 110$\pm$99 & $3.54^{+0.06\:(0.05)}_{-0.06\:(0.05)}$ & 0.91$\pm$0.06 \\
1315 & 10928043 & 13.137 & 6415$\pm$88 & 120$\pm$45 & $3.69^{+0.05\:(0.04)}_{-0.05\:(0.02)}$ & 0.70$\pm$0.05 \\
1344 & 4136466 & 13.446 & 6038$\pm$70 & 112$\pm$41 & $3.52^{+0.05\:(0.03)}_{-0.05\:(0.03)}$ & 1.07$\pm$0.05 \\
1379 & 7211221 & 13.687 & 5870$\pm$70 & 51$\pm$29 & $3.63^{+0.04\:(0.02)}_{-0.04\:(0.02)}$ & 0.95$\pm$0.04 \\
1442 & 11600889 & 12.521 & 5549$\pm$93 & 394$\pm$220 & $4.00^{+0.11\:(0.10)}_{-0.12\:(0.11)}$ & 0.41$\pm$0.11 \\
1445 & 11336883 & 12.320 & 6529$\pm$80 & 102$\pm$27 & $4.09^{+0.04\:(0.01)}_{-0.04\:(0.02)}$ & 0.26$\pm$0.04 \\
1474 & 12365184 & 13.005 & 6743$\pm$113 & 286$\pm$19 & $3.92^{+0.04\:(0.02)}_{-0.04\:(0.02)}$ & 0.35$\pm$0.04 \\
1478 & 12403119 & 12.450 & 5697$\pm$60 & 279$\pm$108 & $3.92^{+0.07\:(0.05)}_{-0.07\:(0.06)}$ & 0.50$\pm$0.07 \\
1525 & 7869917 & 12.082 & 6905$\pm$87 & 260$\pm$50 & $4.23^{+0.05\:(0.03)}_{-0.05\:(0.03)}$ & 0.02$\pm$0.05 \\
1529 & 9821454 & 14.307 & 6314$\pm$89 & 74$\pm$58 & $3.60^{+0.06\:(0.04)}_{-0.06\:(0.04)}$ & 0.91$\pm$0.06 \\
1530 & 11954842 & 13.029 & 6266$\pm$74 & 51$\pm$29 & $3.61^{+0.04\:(0.01)}_{-0.04\:(0.02)}$ & 0.81$\pm$0.04 \\
1531 & 11764462 & 13.069 & 6069$\pm$78 & 553$\pm$169 & $3.51^{+0.10\:(0.10)}_{-0.04\:(0.02)}$ & 0.89$\pm$0.07 \\
1532 & 11656246 & 12.841 & 6449$\pm$97 & 105$\pm$54 & $3.71^{+0.05\:(0.03)}_{-0.05\:(0.03)}$ & 0.61$\pm$0.05 \\
1534 & 4741126 & 13.470 & 6401$\pm$93 & 65$\pm$43 & $3.63^{+0.04\:(0.02)}_{-0.05\:(0.03)}$ & 0.79$\pm$0.05 \\
1535 & 11669125 & 13.046 & 6190$\pm$78 & 163$\pm$86 & $3.69^{+0.06\:(0.04)}_{-0.06\:(0.05)}$ & 0.65$\pm$0.06 \\
1536 & 12159249 & 12.710 & 6059$\pm$94 & 256$\pm$123 & $3.71^{+0.10\:(0.09)}_{-0.08\:(0.07)}$ & 0.67$\pm$0.09 \\
1573 & 5031857 & 14.373 & 6105$\pm$97 & 173$\pm$103 & $3.52^{+0.08\:(0.07)}_{-0.05\:(0.02)}$ & 1.01$\pm$0.06 \\
1667 & 5015913 & 12.989 & 5692$\pm$101 & 123$\pm$35 & $3.66^{+0.05\:(0.03)}_{-0.05\:(0.02)}$ & \nodata \\
1740 & 6762829 & 13.941 & 5901$\pm$83 & 409$\pm$110 & $3.54^{+0.08\:(0.07)}_{-0.06\:(0.04)}$ & \nodata \\
1814 & 5621125 & 12.538 & 7062$\pm$105 & 236$\pm$95 & $3.88^{+0.06\:(0.05)}_{-0.06\:(0.04)}$ & 0.41$\pm$0.06 \\
1822 & 5124667 & 12.443 & 6222$\pm$80 & 160$\pm$49 & $4.12^{+0.05\:(0.02)}_{-0.05\:(0.02)}$ & 0.09$\pm$0.05 \\
1886 & 9549648 & 12.239 & 6346$\pm$91 & 188$\pm$49 & $3.94^{+0.05\:(0.02)}_{-0.05\:(0.03)}$ & 0.04$\pm$0.05 \\
1909 & 10130039 & 12.776 & 6094$\pm$72 & 483$\pm$137 & $3.89^{+0.09\:(0.08)}_{-0.09\:(0.08)}$ & 0.52$\pm$0.09 \\
1952 & 7747425 & 14.601 & 6000$\pm$100 & 143$\pm$75 & $3.50^{+0.06\:(0.05)}_{-0.04\:(0.01)}$ & 0.92$\pm$0.05 \\
1964 & 7887791 & 10.687 & 5543$\pm$60 & 950$\pm$504 & $4.21^{+0.25\:(0.24)}_{-0.25\:(0.25)}$ & 0.22$\pm$0.25 \\
2008 & 8098728 & 10.800 & 6665$\pm$65 & 281$\pm$93 & $4.21^{+0.06\:(0.04)}_{-0.08\:(0.07)}$ & \nodata \\
2025 & 4636578 & 13.781 & 6234$\pm$100 & 186$\pm$98 & $3.68^{+0.07\:(0.05)}_{-0.07\:(0.05)}$ & 0.58$\pm$0.07 \\
2027 & 8556077 & 11.826 & 6649$\pm$78 & 47$\pm$27 & $4.21^{+0.04\:(0.02)}_{-0.04\:(0.01)}$ & \nodata \\
2075 & 10857519 & 12.217 & 6422$\pm$77 & 394$\pm$84 & $4.12^{+0.06\:(0.05)}_{-0.06\:(0.05)}$ & 0.03$\pm$0.06 \\
2086 & 6768394 & 13.959 & 6180$\pm$103 & 90$\pm$54 & $3.73^{+0.05\:(0.03)}_{-0.05\:(0.03)}$ & 0.40$\pm$0.05 \\
2087 & 6922710 & 11.863 & 6223$\pm$109 & 755$\pm$331 & $3.79^{+0.18\:(0.17)}_{-0.17\:(0.16)}$ & 0.48$\pm$0.17 \\
2110 & 11460462 & 12.189 & 6470$\pm$112 & 123$\pm$32 & $4.24^{+0.04\:(0.02)}_{-0.04\:(0.02)}$ & 0.08$\pm$0.04 \\
2148 & 6021193 & 13.353 & 5604$\pm$123 & 284$\pm$140 & $3.58^{+0.08\:(0.07)}_{-0.09\:(0.08)}$ & 0.33$\pm$0.09 \\
2149 & 10617017 & 12.071 & 6314$\pm$73 & 128$\pm$42 & $4.19^{+0.05\:(0.02)}_{-0.04\:(0.02)}$ & $-$0.02$\pm$0.05 \\
2178 & 2014991 & 12.396 & 6420$\pm$101 & 299$\pm$121 & $3.83^{+0.09\:(0.09)}_{-0.08\:(0.07)}$ & \nodata \\
2230 & 8914779 & 11.511 & 6266$\pm$84 & 163$\pm$55 & $4.14^{+0.05\:(0.03)}_{-0.06\:(0.04)}$ & \nodata \\
2249 & 4761060 & 12.301 & 6756$\pm$107 & 259$\pm$61 & $3.92^{+0.06\:(0.05)}_{-0.06\:(0.04)}$ & \nodata \\
2295 & 4049901 & 11.671 & 5453$\pm$93 & 46$\pm$32 & $4.19^{+0.04\:(0.01)}_{-0.04\:(0.01)}$ & 0.29$\pm$0.04 \\
2414 & 8611832 & 13.584 & 5889$\pm$74 & 54$\pm$32 & $3.87^{+0.04\:(0.02)}_{-0.04\:(0.02)}$ & 0.64$\pm$0.04 \\
2595 & 8883329 & 13.223 & 6741$\pm$115 & 95$\pm$42 & $3.88^{+0.05\:(0.02)}_{-0.05\:(0.02)}$ & 0.23$\pm$0.05 \\
2612 & 9602613 & 11.830 & 5450$\pm$59 & 56$\pm$24 & $4.31^{+0.04\:(0.01)}_{-0.04\:(0.01)}$ & \nodata \\
5145 & 5263802 & 11.494 & 6916$\pm$105 & 428$\pm$95 & $4.14^{+0.06\:(0.04)}_{-0.05\:(0.04)}$ & \nodata \\
5814 & 10616656 & 12.138 & 6506$\pm$85 & 233$\pm$202 & $4.19^{+0.10\:(0.09)}_{-0.10\:(0.09)}$ & \nodata \\
\enddata
\tablenotetext{a}{The top eight entries had no assigned KOI identifier at the time of writing.}
\tablenotetext{b}{Effective temperatures are from \citet[][]{Pinsonneault}.}
\tablenotetext{c}{Uncertainties in brackets are produced by Eq.~(\ref{eq:gscaling}) and do not take into account the adopted figure of $0.04\:{\rm dex}$ for the accuracy.}
\end{deluxetable}


\end{document}